\documentclass[draft]{agujournal2019}
\usepackage{url}
\usepackage{lineno}
\usepackage[inline]{trackchanges}
\usepackage{soul}
\usepackage{bm}
\usepackage{amsmath}
\usepackage{setspace}
\usepackage{caption}
\usepackage{xr}
\journalname{Journal of Geophysical Research: Solid Earth}

\begin{document}

\newcommand{\ib}[1]{\textbf{\textit{#1}}}
\newcommand{\norm}[1]{\left\lVert{#1}\right\rVert}

\def\fm{3D-1400}
\def\fmm{3D-500A}

\title{Instantaneous physics-based ground motion maps using reduced-order modeling}

\authors{John M. Rekoske\affil{1}, Alice-Agnes Gabriel\affil{1,2}, Dave A. May\affil{1}}

\affiliation{1}{Scripps Institution of Oceanography, University of California San Diego, La Jolla, USA}
\affiliation{2}{Department of Earth and Environmental Sciences, Ludwig-Maximillians-Universität München, Munich, Germany}
\correspondingauthor{John M. Rekoske}{jrekoske@ucsd.edu}

\begin{keypoints}
    \item Physics-based simulations are useful for seismic hazard assessment and ground motion prediction but are computationally expensive.
    \item Our reduced-order model accurately predicts simulated peak ground velocities for scenarios, including 3D wavefield and topographic effects.
    \item The reduced-order model produces physics-based ground motion maps in milliseconds for variable source depths and focal mechanisms.
\end{keypoints}

\begin{abstract}
Physics-based simulations of earthquake ground motion are useful to complement recorded ground motions.
However, the computational expense of performing numerical simulations hinders their applicability to tasks that require real-time solutions or ensembles of solutions for different earthquake sources.
To enable rapid physics-based solutions, we present a reduced-order modeling (ROM) approach based on interpolated proper orthogonal decomposition (POD) to predict peak ground velocities (PGVs).
As a demonstrator, we consider PGVs from regional 3D wave propagation simulations at the location of the 2008 Mw 5.4 Chino Hills earthquake using double-couple sources with varying depth and focal mechanisms.
These simulations resolve frequencies $\leq$ 1.0 Hz and include topography, viscoelastic attenuation, and S-wave speeds $\geq$ 500 m/s.
We evaluate the accuracy of the interpolated POD ROM as a function of the approximation method.
Comparing the radial basis function (RBF), multilayer perceptron neural network, random forest, and $k$-nearest neighbor, we find that the RBF interpolation gives the lowest error ($\approx$ 0.1 cm/s) when tested against an independent dataset.
We also find that evaluating the ROM is $10^7-10^8$ times faster than the wave propagation simulations.
We use the ROM to generate PGV maps for one million different focal mechanisms, in which we identify potentially damaging ground motions and quantify correlations between focal mechanism, depth, and accuracy of the predicted PGV.
Our results demonstrate that the ROM can rapidly and accurately approximate the PGV from wave propagation simulations with variable source properties, topography, and complex subsurface structure.
\end{abstract}

\section*{Plain Language Summary}
Computer simulations can be used to predict
the intensity of ground shaking caused by earthquakes. However,
these simulations require significant amounts of computing time
to obtain accurate results, making them too slow for real-time systems.
In this work, we apply a technique that allows us to obtain
instantaneous and accurate approximations of
a scenario earthquake simulation using simplified models.
The simplified models are fast enough to evaluate
this earthquake scenario on demand.
We use the simplified models to predict shaking
intensities for one million different
fault planes at the location of the 2008 Chino Hills earthquake.
By analyzing the predicted shaking intensities
from these earthquakes, we identify possible areas of strong
shaking, areas that might otherwise be unidentified
if fewer earthquake scenarios were examined.
These results suggest that these types of simplified models
could be considered to assess regional earthquake hazards and to
inform shaking intensity predictions in earthquake early warning.

\section{Introduction}
A fundamental challenge in seismology and earthquake engineering is the estimation of the ground shaking caused by earthquakes, %
especially for areas with high population densities or containing
critical infrastructure and in the near-source region where observations are sparse.
For example, after large earthquakes, it is important to rapidly characterize the shaking intensity
to identify areas of potential damage and guide
rapid response efforts. Accurately estimating shaking intensities in %
near-real-time is one of the primary goals of the
ShakeMap product developed by the U.S. Geological Survey \cite{wordenShakeMapManualShakeMap2016}.
Additionally, real-time estimates of the expected ground motion intensity
are needed in earthquake early warning (EEW) to issue alerts to people
and critical systems, ideally in advance of the strongest shaking \cite{kohlerEarthquakeEarlyWarning2017}.
Also, systematically quantifying the potential level of ground shaking from as of
yet unobserved earthquakes that may someday occur is important for earthquake preparedness
and seismic hazard assessment \cite<e.g.,>[]{gravesCyberShakePhysicsbasedSeismic2011,Baker2014,bommerSelectionGroundmotionPrediction2010}.

Several strategies have been developed to estimate earthquake ground motions. %
Instrumental recordings and non-instrumental data such as ``Did You Feel It?'' reports \cite{waldUSGSDidYou2011}
directly inform the amplitudes and spatial distribution of seismic ground motion. %
However, the generally sparse spatial coverage of observations leads to significant uncertainties
in the shaking distributions \cite{waldQuantifyingQualifyingUSGS2008}. %
To fill in the gaps, recent approaches include spatial and spectral interpolation
\cite{wordenSpatialSpectralInterpolation2018a}, Gaussian process regression \cite{tamhidiConditionedSimulationGround2021},
and conditional simulation \cite{engler2022partitioning,bailey2022adapting}, but rapidly estimating the full distribution of
shaking intensity remains a challenge. Most EEW systems and
large-scale seismic hazard studies
\cite<e.g.,>[]{stirlingNationalSeismicHazard2012,kohlerEarthquakeEarlyWarning2017,petersen2018UpdateUS2020,
melettiNewItalianSeismic2021} instead use empirical ground motion models (GMMs) which %
give instantaneous ground motion estimates for given source, path, and site
parameters. GMMs rely on the availability of a large ground motion dataset recorded
in a specific region or tectonic setting \cite<e.g.,>[]{bozorgniaNGAWest2ResearchProject2014,bozorgniaNGASubductionResearchProgram2021,gouletNGAEastGroundMotionCharacterization2021},
which may not always be available.
Thus, GMMs typically rely on the ergodic assumption \cite{andersonProbabilisticSeismicHazard1999}
where all the data from a certain tectonic environment are grouped together
to develop a single model that describes the average scaling
with magnitude, distance, and site conditions.
However, the uncertainty of GMM predictions can be reduced
with nonergodic source, path, or site correction terms if they are well-constrained %
\cite<e.g.,>[]{baltayUncertaintyVariabilityEarthquake2017,sahakianGroundMotionResiduals2019,parkerEmpiricalMapBasedNonergodic2022}.
Going beyond the source-based approach of GMMs, other
methods have been developed that directly use ground motion or wavefield
observations to forecast the shaking at distant locations
\cite{hoshiba2015numerical,kodera2018propagation,furumura2019early,cochran2019event,oba2020data}.
A recent wavefield-based approach developed by \citeA{nagata2023seismic} uses
reduced-order modeling to reconstruct seismic wavefields from sparse
measurements, although their approach does not include the source parameterization
that is needed to solve inverse problems.

Complementary to GMM and wavefield-based
prediction techniques, numerical simulations of earthquake rupture and seismic %
wave propagation provide physics-informed estimates
of shaking intensity \cite<e.g.,>[]{aagaardCharacterizationNearsourceGround2001,
olsenStrongShakingAngeles2006,moschetti3DGroundmotionSimulations2017,
rodgersRegionalScale3D2020,taufiqurrahmanBroadbandDynamicRupture2022a}. These simulations
can assume a kinematic description of the earthquake source %
\cite<e.g.,>[]{gravesKinematicGroundmotionSimulations2016},
or solve for the earthquake slip history with spontaneous dynamic
rupture models \cite<e.g.,>[]{rippergerVariabilityNearfieldGround2008,withersValidationDeterministicBroadband2018}.
In both cases, the seismic wave equation is solved to
simulate the propagation of seismic
waves within the Earth. Physics-based 3D ground motion simulations have advanced
in terms of realism and agreement with real data due to increasingly dense observations,
a better understanding of Earth structure and ever-increasing computing power.
For example, recent advances in community velocity model (CVM) development
in southern California have been used to inform physics-based
seismic hazard assessment \cite{gravesCyberShakePhysicsbasedSeismic2011,milnerPhysicsbasedNonergodicPSHA2021}.
Important progress has been made in generating broadband synthetics
which usually involve a stochastic component for the high frequencies
\cite<e.g.,>[]{booreStochasticSimulationHighfrequency1983,frankelBroadbandSyntheticSeismograms2018}.

Deterministically resolving the high-frequency wavefield makes physics-based ground motion modeling
 computationally expensive,
typically requiring high-performance computing (HPC) technology \cite<e.g.,>[]{Cui2013,Heinecke2014,rodgersRegionalScale3D2020,Savran2020,pitarkaDeterministic3DGroundMotion2022b}.
The high cost of each high-resolution numerical simulation
challenges applications in which many forward models must be evaluated, such as dynamic earthquake source inversion problems
\cite<e.g.,>[]{gallovicBayesianDynamicFinitefault2019b} or physics-based probabilistic seismic hazard assessment.
Even though modeling simplifications help reduce the computational expense (e.g., by setting an upper threshold for the minimum S-wave velocity, $V_S$),
regional-scale broadband simulations can require up to $10^3-10^4$
CPU hours (CPUh) of computing time \cite<e.g.,>[]{huHzDeterministic3D2022,taufiqurrahmanBroadbandDynamicRupture2022a}.

To address these challenges, recent efforts
have enabled the rapid generation of synthetic seismograms using Green's functions or deep learning. %
For global seismology, 1.0 Hz discrete Green's functions for axisymmetric Earth models are available
on-demand via InstaSeis \cite{vandrielInstaseisInstantGlobal2015}, and
3D Green's functions based on the spectral element method \cite{komatitschIntroductionSpectralElement1999a}
are typically available a few hours after $M_w>5.5$ earthquakes \cite{trompRealtimeSimulationsGlobal2010}.
However, the limited resolution of global approaches are not suitable
for regional ground motion applications.
At the regional scale, \citeA{wangMoving1D3D2020} created a 0.5 Hz 3D Green's function database for stations in the southern California seismic network
to enable real-time moment tensor inversions with 3D models. However, such Green's function
databases often require several terabytes of storage, presenting
a major challenge in accessibility.
Other approaches used deep learning to learn approximate solutions to the wave equation and produce
seismograms for arbitrary velocity models \cite{moseleyFastApproximateSimulation2018,
moseleyDeepLearningFast2020,smithEikonetSolvingEikonal2020,yangSeismicWavePropagation2021a,yang2023rapid,
rasht-beheshtPhysicsInformedNeuralNetworks2022}
or create neural network GMMs based on simulations \cite{withersMachineLearningApproach2020a}.
However, machine-learning approaches have not yet directly reproduced
3D path or topographic effects that are important for ground motion prediction.

Here, we present a surrogate modeling approach
that {aims to address} the need for rapid generation of 3D physics-based synthetics
and ground motion intensity maps. %
Our approach is based on a reduced-order
model (ROM), a particular type of surrogate model. %
In the model order reduction
framework, a high-fidelity forward model can be approximated by a surrogate model,
which typically has many fewer degrees of freedom and is fast to evaluate.
As a result, a well-trained ROM can enable real-time
and many-query evaluation of a system and
can closely approximate the solution of an expensive forward simulation.
We predict the peak-ground velocity (PGV) from
a scenario of regional ground motion simulation.
Our method %
produces physics-informed estimates instantaneously, thus, at the same speed as GMMs. %
 While ROMs have rarely been
applied in geophysics thus far, they have been demonstrated %
to accurately approximate solutions to linear and non-linear physics-based models in other fields of science and engineering
\cite<e.g.,>[]{willcoxBalancedModelReduction2002,bennerSurveyProjectionbasedModel2015,hesthavenReducedBasisMethods2022}.

The structure of this paper is organized as follows. We first provide an overview of our
ROM development workflow based on an interpolated proper orthogonal decomposition (iPOD) technique in
Section \ref{sec:overview}.
We use simulations to create simulated PGV maps for different earthquake sources and decompose
them using iPOD. We create the ROM by fitting function approximators that predict the iPOD coefficients
for different values of the input source parameters.
We can then use the ROM to instantaneously predict PGV maps for new earthquakes that were not modeled.
We describe the details of the
forward problems for both a 1D layered and a 3D velocity model with topography in
Section \ref{sec:forward}. We describe
how we will use four different function approximators to predict the simulation output %
in Section \ref{sec:rom}. In our results Section \ref{sec:results}, we evaluate the
 accuracy of the ROMs using an independent
testing dataset. We provide illustrative examples %
highlighting the ROM's ability to reproduce PGV maps faithfully. Finally,
we present a potential application and sensitivity analysis
emphasizing the importance of hypocentral depth in PGV predictions.
Lastly, we discuss in Section \ref{sec:discuss} how reduced-order modeling may path new avenues for physics-based hazard analysis %
and for using high-performance computing empowered simulations in real-time systems by enabling the instantaneous generation
of physics-informed ground motion maps. %
\section{Methods}
\label{sec:methods}

\subsection{Reduced-order modeling workflow}
\label{sec:overview}
We consider the earthquake source to be a point source in space, parameterized
by its hypocentral depth, strike, dip, and rake.
These four parameters will be denoted by $\bm{p}$.
The ground motion intensity metric we use is the surface peak-ground velocity (PGV)
denoted by $\bm q$.
We define the mapping between the parameters $\bm p$ and the PGV $\bm q$
using a physics-based approach given by the
(visco-)elastic wave equation for an isotropic medium.
When a 3D velocity model is used to describe the Earth's crustal
structure and/or non-planar topography is considered, this problem must be solved numerically.
The computational cost of performing these numerical wave propagation simulations
precludes their direct usage within
real-time or rapid response infrastructure,
as well as many-query tasks, for example as required for
global sensitivity analysis and forward uncertainty quantification.

\begin{figure}
    \centering
    \includegraphics[width=\textwidth]{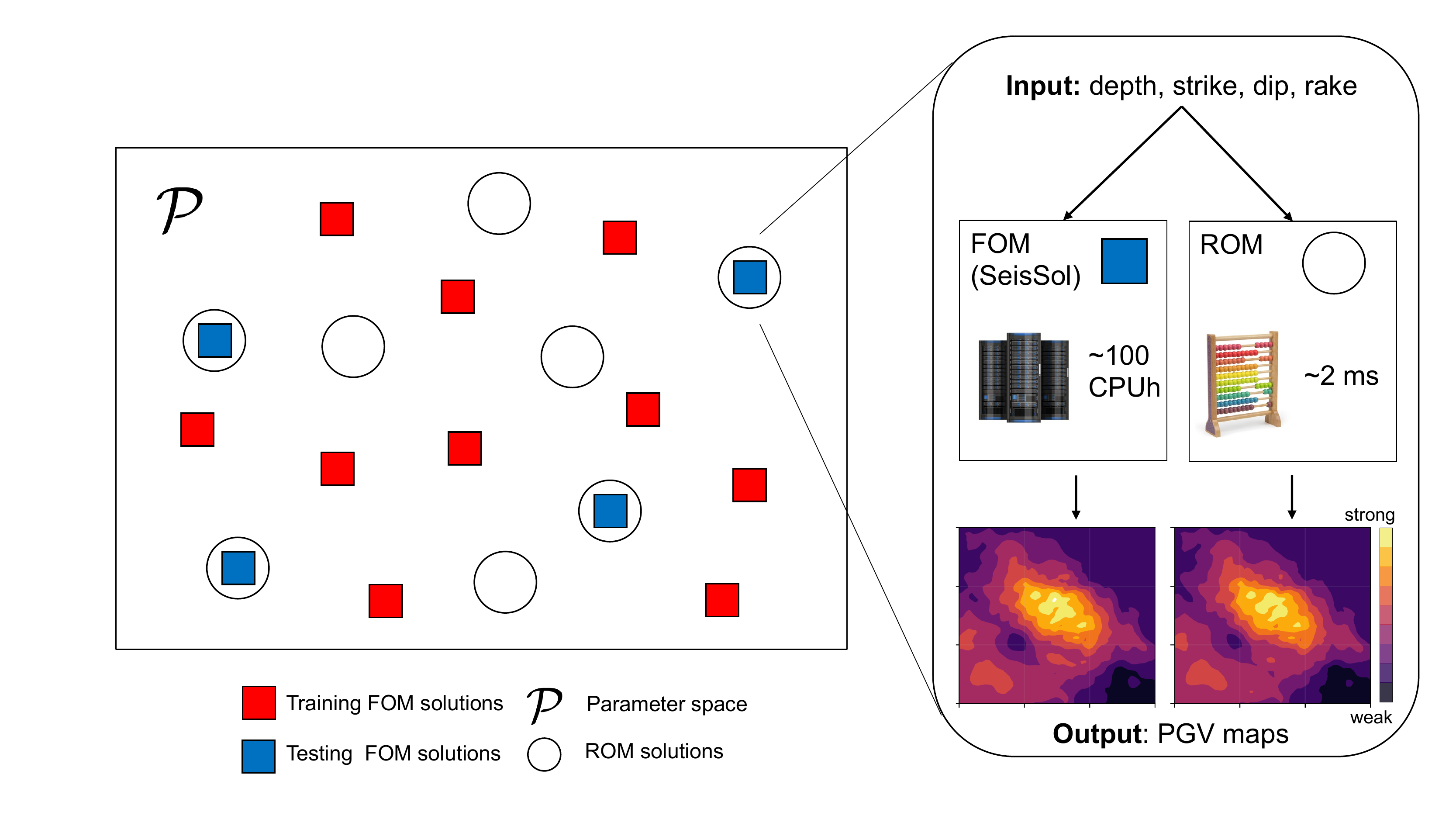}
    \caption{Conceptual illustration of the reduced-order model (ROM) development %
    and validation procedures. We first obtain full-order model (FOM) solutions for points in the parameter
    space $\mathcal{P}$ and divide them into training (red squares)
    and testing solutions (blue squares).
    Here, FOM refers to the forward problem of a numerical solution of the wave equation and producing a surface peak ground velocity (PGV) map.
    After training the ROM using the
    training FOM solutions, one can evaluate the ROM for any
    new point in $\mathcal{P}$ (open circles). The ROM solutions are validated against
    the testing FOM solutions (overlapping circles and blue squares).
    As shown on the right, the FOM and ROM output similar PGV maps that
    depend on the input source parameters; the ROM, however, is
    a computationally inexpensive model with fewer degrees of freedom.}
    \label{fig:basic}
\end{figure}

Referring to the numerical method used to solve the wave equation and produce the PGV $\bm q$
as the full-order model (FOM),
we wish to construct a reduced-order model (ROM)
that approximates $\bm q$ given the earthquake parameters $\bm p$ as input (Figure \ref{fig:basic}).
While solving the FOM is a computationally
expensive problem that requires HPC, the ROM
has fewer degrees of freedom %
resulting in lower computational complexity compared to the FOM.
Thus, the ROM is much faster to evaluate than the FOM.
To inform the ROM, we obtain FOM solutions for a set of points
$\{\bm{p}_1$, $\bm{p}_2$, $\ldots$, $\bm{p}_{n_s}\}$
where $n_s$ is the total number of simulations
that will be performed. Each parameter vector $\bm{p}_i$ is contained
within the four-dimensional space $\mathcal{P}$ which encompasses
the range of source depths and focal mechanism angles
we consider; the ROM is only applicable for sources
contained within $\mathcal{P}$.
We then split the set of FOM solutions into two groups: training
FOM solutions and testing FOM solutions. After using the training FOM
solutions to train the ROM, we validate it against the testing FOM
solutions. If there is a good agreement, then the ROM can be used
as an efficient substitute, or surrogate, for the FOM.
\subsection{Forward model}
\label{sec:forward}
Our forward problem is the simulation
of PGV maps that result from earthquakes with a given
hypocentral depth and focal mechanism.
We numerically simulate the 3D propagation of seismic
waves produced by each source, varying
the strike, dip, rake, and hypocentral depth of the source in each simulation.
Our source parameterization is motivated by seismicity in Southern California \cite{SCEDC2013,field2014uniform} but simplified for the proof-of-concept scope of this study.
As a source-time function that describes the
rate of earthquake moment release, we use
$\dot{M}(t) = \frac{M_0 t}{T^2}e^{-t/T}$ with $T=0.4$ s.
This results in a Brune-type source spectrum with a stress drop
of approximately 5.0 MPa \cite{bruneTectonicStressSpectra1970}.
We do not vary the total scalar seismic moment, prescribing $M_w=5.4$
in all simulations. We assume point-source, double-couple moment tensor earthquake source descriptions,
which is a reasonable assumption for most small earthquakes resulting from shear faulting
\cite{julian1998non}.
The hypocentral depths vary from 2.0 to 20.0 km.

\begin{figure}
    \centering
    \includegraphics[width=\textwidth]{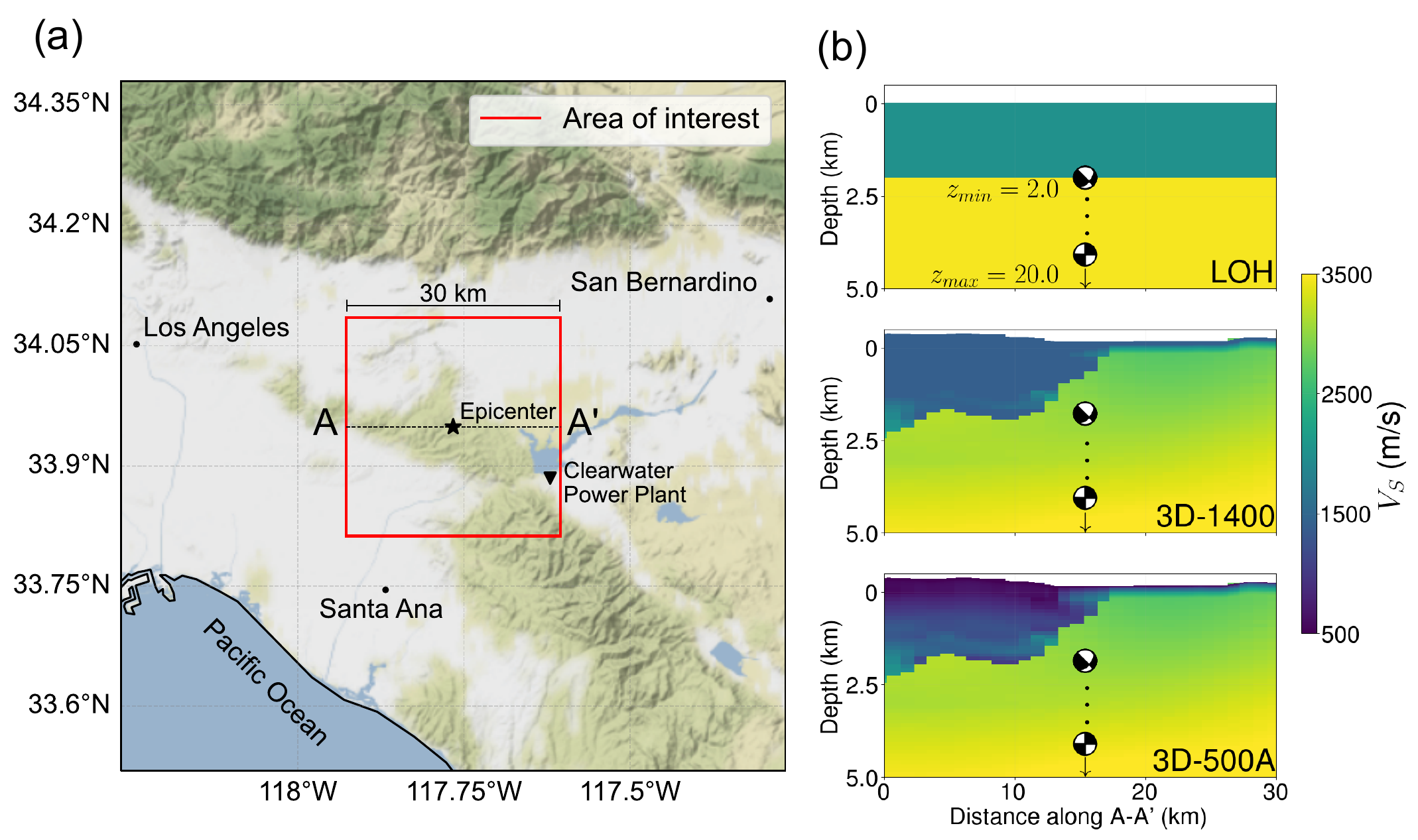}
    \caption{(a) Map of the area of interest (red box) and the surrounding region in Southern California.
        The epicenter of the simulated earthquakes (which is the
        same as the 2008 $M_w$ 5.4 Chino Hills earthquake) is indicated by a black
        star and a site of interest is indicated by a black triangle.
        (b) Cross-sections (A - A') of the velocity models used for
        wave propagation simulation of the three forward models
        (LOH, 3D-1400, and 3D-500A). Note that the cross-sections
        correspond to the area of interest, though our simulations
        use larger meshes to prevent any potential spurious boundary reflections.
        As shown in the cross-sections, we vary the source depth and focal mechanism for
        each of the $n_s=5000$ simulations. The source depths range
        between $z_{min}$=2.0 km and $z_{max}$=20.0 km, though the cross-sections
        show a maximum depth of 5.0 km to better show the near-surface
        velocity structures.}
    \label{fig:forward}
\end{figure}

To generate ground motion maps from the seismic wave simulations for each source, we first define a high-resolution
area of interest. We select a 30~km by 30~km square area
(Figure \ref{fig:forward}a) centered at the epicenter of
the 2008 $M_w$ 5.4 Chino Hills earthquake (33.949$^\circ$ N, 117.776$^\circ$ W), an
area that has potentially significant topographic and basin effects on
ground motion \cite{tabordaGroundmotionSimulationValidation2013}. Note that the complete computational mesh is larger (150 $\times$ 150
$\times$ 100 km) than the area of interest to prevent artificial boundary reflections
from affecting the PGV measurements. %
We limit the computational expense
of the forward simulations by selecting $f_{max}=1.0$ Hz as the maximum
frequency to fully resolve.

For each earthquake source, we perform forward simulations with three different
setups of increasing complexity (Figure \ref{fig:forward}b).
The first setup uses a flat, layer-over-halfspace velocity
model, which defines a 1D velocity structure
(denoted as ``LOH'', see Table S1 for the
material parameters). The second setup ``\fm'' uses a 3D community velocity model for Southern California, SCEC CVM-H
v15.1.0 \cite{shawUnifiedStructuralRepresentation2015},
and includes topography. ``1400'' indicates that we set a floor for the
minimum S-wave velocity at $V_{S,\text{min}}=1400$ m/s. In both
LOH and \fm, we assume linear, elastic wave propagation.
The third setup ``\fmm'' is similar to
\fm \ except that $V_{S,\text{min}}=500$ m/s and that we account for
frequency-independent viscoelastic attenuation with
$Q_p=0.1V_S$ and $Q_s=0.05V_S$ (for $V_S$ measured in m/s) \cite{olsen2003estimation}.
The viscoelastic attenuation setup assumes a Maxwell material with three damping mechanisms
with their respective relaxation frequencies of 0.05, 0.5, and 5.0 Hz
\cite{uphoff2017extreme}. These different modeling setups allow us to examine
how the ROM predictions and accuracies are affected by the modeling assumptions related to each simulation regime.

To perform the seismic wave propagation simulations for each source,
we use SeisSol (\url{https://github.com/SeisSol}), an open-source code based on the arbitrary high-order accurate derivative
discontinuous Galerkin (ADER-DG) method \cite{dumbserArbitraryHighorderDiscontinuous2006} with
end-to-end optimization for HPC infrastructure \cite<e.g.,>[]{Rettenberger2016,Uphoff2017,Krenz2021}.
SeisSol uses fully non-uniform unstructured tetrahedral meshes that can statically adapt to complex 3D subsurface structures and topography.
We create 3D tetrahedral meshes for the simulations using
velocity-aware meshing, i.e., automatically adapting SeisSol's mesh resolution to the 1D or 3D velocity model (Figure \ref{fig:forward}b).
We ensure sufficient resolution using four elements with basis functions of polynomial degree of 4 per shortest
wavelength \cite{Kaeser2008}. To inform the 3D models, we obtain the
CVM-H velocity model using the SCEC UCVM software \cite{smallSCECUnifiedCommunity2017a}.
We obtain the topography data from the 1-meter resolution digital
elevation model from the U.S. Geological Survey 3D Elevation Program (see
data availability statement). We create the meshes with Gmsh \cite{geuzaineGmsh3DFinite2009},
resulting in 550,001 tetrahedral elements for LOH, 129,888 elements
for 3D-1400, and 1,189,840 elements for 3D-500A.
All simulations were performed
using a degree four polynomial in space and time
(translating to a fifth-order accurate scheme in space and time)
and were executed on SuperMUC-NG using
17 compute nodes. Each compute node consists of
48 cores employing Intel Skylake Xeon Platinum 8174 processors.
Using these resources
the runtimes per simulation were 59.5 s, 23.0 s, and 390.0 s for the
LOH, 3D-1400, and 3D-500A models, respectively.
Assuming perfect utilization of the 17$\times48$ cores, these runtimes translate to
13.5, 5.2, and 88.4 CPUh.
We save the three-component free-surface velocity time-series
for $n_r=3600$ receivers with a spacing of 500 m and sampling rate
of 10.0 Hz. To create a ground motion map for each earthquake source, the last
step is to compute the PGV from the time series of the horizontal
components. We demean, detrend, and apply a 1.0 Hz lowpass filter
to the time series before combining the two horizontal components
using the orientation-independent RotD50 measure \cite{booreOrientationindependentNongeometricmeanMeasures2010}.

\subsection{Reduced-order model}
\label{sec:rom}
We now describe in detail how to define the ROM that approximates
the mapping  between the parameters $\bm p$ and the PGV map $\bm q$.
\subsubsection{Parameter space sampling}
To inform the ROM, we first select the points in the parameter space
$\mathcal{P}$ where we will obtain the FOM solutions. To determine these
points, we use a low-discrepancy, four-dimensional Halton sequence
\cite{haltonEfficiencyCertainQuasirandom1960}.
The low-discrepancy property of this sequence means that it closely
approximates points drawn from a uniform distribution, giving us points
that evenly cover our entire parameter space.
We use the first $n_s=5000$ points
from this sequence, where $n_s$ is the total number of earthquake sources
for which we evaluate the FOM.
We gather the parameters determined by the Halton sequence into
a matrix $\bm{P}=[\ib{p}_1, \ib{p}_2, \ldots ,\ib{p}_{n_s}]$
and the PGV data generated by the FOM into a matrix
$\bm{Q}=[\ib{q}_1, \ib{q}_2, \ldots ,\ib{q}_{n_s}]$.
Generating the simulation data matrix $\bm{Q}$ for $n_s=5000$ sources
requires a total of 67.5 kCPUh, 26 kCPUh, and 442 kCPUh for the
LOH, \fm\, and \fmm\ forward models, respectively.
To create independent datasets for training and testing of the ROM, we
randomly separate the PGV maps (stacked in $\bm Q$) that result from these sources into
training $(\bm Q_\text{train}, \bm P_\text{train})$ and testing $(\bm Q_\text{test}, \bm P_\text{test})$ datasets, where the testing dataset
contains 10\% of the simulated PGV maps ($n_{\text{train}}=4500$,
$n_{\text{test}}=500)$. The accuracy of the trained ROM can then be evaluated
using PGV maps from the independent testing dataset.

As the goal of the ROM is to interpolate the known PGV maps
for new sources in $\mathcal{P}$ that we did not simulate, we
measure the Euclidean distances between each testing parameter
$\bm{p} \in \bm{P}_{\text{test}}$ and
its nearest-neighboring training parameter $\bm{p_{\text{nearest}}} \in \bm{P}_{\text{train}}$, i.e.,
\begin{equation*}
    d_{\text{nearest}}(\bm{p})=||\bm{p}-\bm{p_{\text{nearest}}}||_2.
\end{equation*}
These are the distances after normalizing
each dimension of the parameter space (depth, strike, dip, and rake)
to range between 0 and 1.
In our testing dataset, the distribution of distances has a mean value
of 0.09 (Figure S1).
We also create a set of sources $\bm{P}_{\text{uniform}}$
that contains $10^6$ uniformly distributed sources that evenly
cover $\mathcal{P}$.
While it is not possible to evaluate the FOM to obtain error estimates for
the $10^6$ points in $\bm{P}_{\text{uniform}}$, we show that our testing distribution
is a good proxy of the errors
as indicated by the mean and standard deviations of the
distances for both $\bm{P}_{\text{test}}$ and
$\bm{P}_{\text{uniform}}$ (Figure S1). For any earthquake
source contained in $\mathcal{P}$, the distance to the nearest training point will be,
on average, 0.08. The maximum possible distance to any training
FOM location is 0.17 units which is
our reported ``fill distance'' \cite{de2010stability} for our sampled FOM locations.

\subsubsection{Interpolated proper orthogonal decomposition (iPOD)}
\begin{figure}
    \centering
    \includegraphics[width=\textwidth]{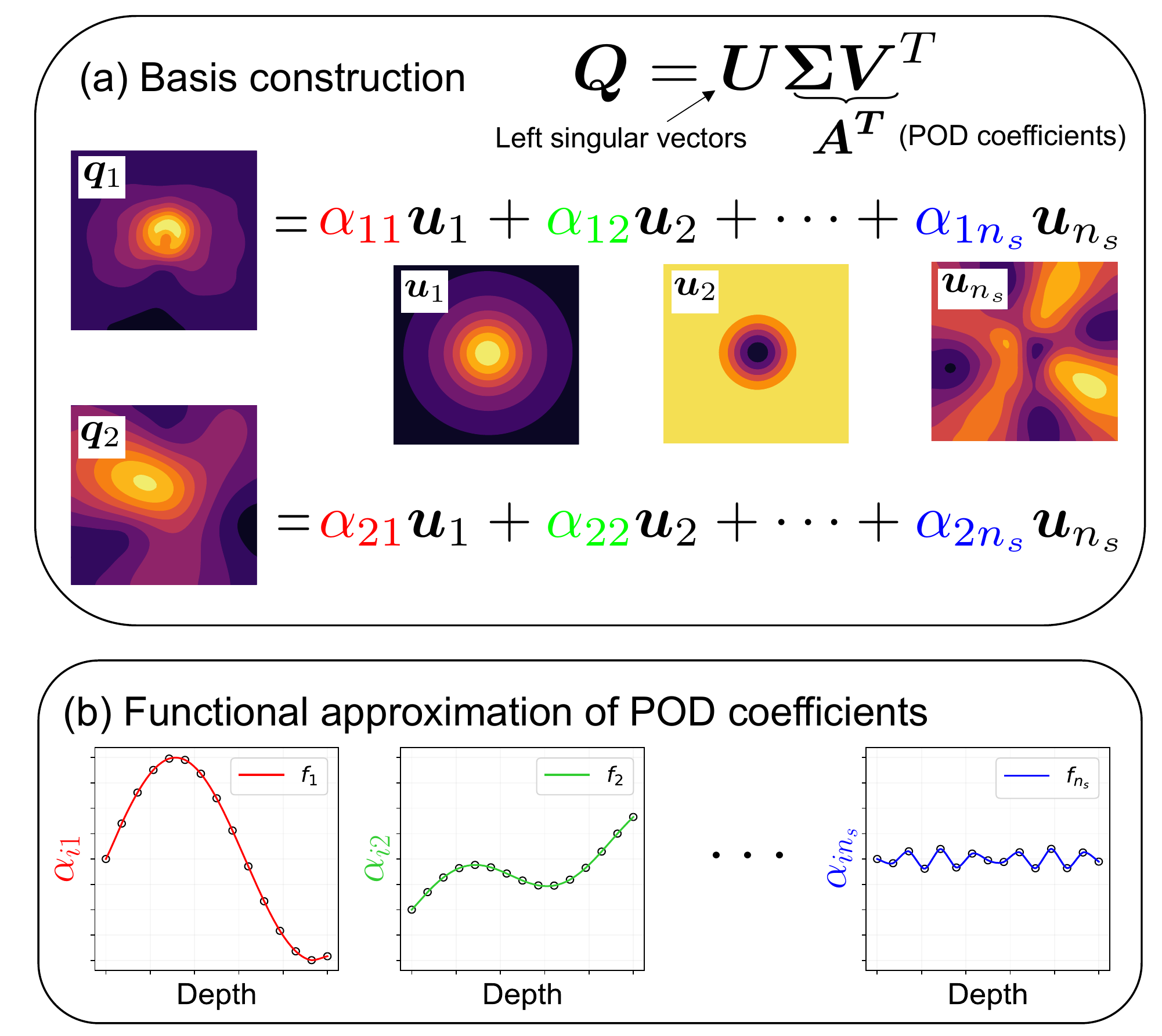}
    \caption{Conceptual illustration of a reduced-order model based on interpolated proper orthogonal decomposition (iPOD).
        We first decompose the data matrix of PGV maps $\bm{Q}$
        using the singular value decomposition (a). The goal is then to
        fit a series of functions $f_j$ (b) that predict the proper orthogonal decomposition (POD) coefficients $\alpha_{ij}$ for
        different values of the input source parameters (e.g., hypocentral depth).
        The ROM-predicted PGV maps are obtained by evaluating the functions to obtain
        predicted POD coefficients, and then performing a weighted sum of the
        left singular vectors $\bm{u}_j$.}
    \label{fig:concept}
\end{figure}

We use a method for defining a ROM called the interpolated proper orthogonal decomposition (iPOD)
\cite<e.g.,>[]{bui2003proper, druaultUseProperOrthogonal2005}. In this
technique (Figure \ref{fig:concept}a), we first compute
a modal decomposition for the collected data using the singular value decomposition (SVD),
\begin{equation}
    \bm{Q}=\bm{U\Sigma V^T},
    \label{eq:svd}
\end{equation}
where $\ib{U} = [\bm u_1, \dots, \bm u_{n_s}]$ contains the left singular vectors (or modes) $\bm{u}_i$,
$\bm \Sigma = \text{diag}[\sigma_1, \dots \sigma_{n_s}]$ is a diagonal matrix
containing the singular values $\sigma_i$, and \ib{V} contains the right singular vectors
\cite{berkoozProperOrthogonalDecomposition1993}.
We then define the matrix of POD coefficients $\bm{A}$ to be,
\begin{align}
    \bm{A}^T&=\bm{\Sigma V^T} \\
    \bm{A}&=\bm{V}\bm{\Sigma},
    \label{eq:weights}
\end{align}
where we denote individual entries of $\bm{A}$ as $\alpha_{ij}$.
Using Eqs.~\eqref{eq:svd} and \eqref{eq:weights} we can
express every $\bm q_i \in \bm Q$ using a weighted sum (Figure \ref{fig:concept}a) of the
left singular vectors, i.e.,
\begin{equation}
    \bm{q}(\bm{p}_i) = \bm q_i =\sum_{j=1}^{n_s}\alpha_{ij}\bm{u}_j.
    \label{eq:recon}
\end{equation}

Eq.~\eqref{eq:recon} is a restatement of the factorization defined in Eq.~\eqref{eq:svd}
and thus can only be used to reconstruct values of $\bm q$ for parameters $\bm p \in \bm P$.
To enable predictions for parameters $\bm{p} \notin \bm{P}$,
the iPOD method utilizes a generalization of Eq.~\eqref{eq:recon}
\begin{equation}
    \bm q(\bm p) \approx \tilde{\bm{q}}(\bm{p})=\sum_{j=1}^{n_s}f_j(\bm{p})\bm{u}_j,
    \label{eq:ipod}
\end{equation}
where each function $f_j(\bm p)$, $j = 1, \dots, n_s$
predicts the
POD coefficient for any parameter $\bm p$
associated with the $j^\text{th}$ mode (Figure \ref{fig:concept}b).
The types of functions $f_j(\cdot)$ that we consider are described in Section \ref{sec:rom}.
After fitting the functions $f_j(\bm p)$ with the POD coefficients given by $\alpha_{ij}$,
the PGV map $\tilde{\bm{q}}$ is
obtained by evaluating Eq.~\eqref{eq:ipod}.
In essence, iPOD builds an orthogonal basis $\bm u_i$ based on simulation data $(\bm Q)$
with which it will use to define a predicted PGV map.
The basis $\bm u_i$ are combined linearly, with the associated weight of each basis
defined by a functional approximation of the coefficients $\bm A$.
The POD coefficients needed to reconstruct the training dataset are then given by
$\bm{A}_\text{train}$.

\subsubsection{Function approximation for $\alpha_{ij}$}
We consider four types of function approximators to model the relationships
between the source parameters $\bm p$ and POD coefficients $\alpha_{ij}$.
These include radial basis functions, neural networks, random forests, and $k$-nearest neighbors.
We summarize these function approximation techniques below.

\begin{enumerate}
\item
Radial basis function (RBF) interpolation is commonly
used as an efficient technique for interpolating high-dimensional, unstructured data \cite{lazzaroRadialBasisFunctions2002}
and has been shown to be successful in other studies that use iPOD
\cite<e.g.,>[]{audouzeReducedorderModelingParameterized2009,xiaoNonintrusiveReducedorderModelling2015}.
For RBF interpolation, the interpolator for the $j$-th mode $f_j(\cdot)$ is given by
\begin{equation}
    f_j(\bm{p})=\sum_{i=1}^{n_{\text{train}}} w_{ij}\varphi(|| \bm{p} -\bm{p}_i ||)+\sum_{k=1}^{|V|} b_{kj}\psi_k(\bm{p})
    \label{eq:rbf}
\end{equation}
where the function $\varphi(\cdot)$ is an RBF, $w_{ij}$ in an entry
of the kernel weight matrix $\bm{W}$, $b_{kj}$ is an entry of the
polynomial coefficient matrix $\bm{B}$, and $\psi_k(\cdot)$ are
monomial terms up to a specified degree $d$ contained in the set $V$
with
\begin{equation}
    V = \Biggl\{p_1^qp_2^rp_3^sp_4^t: \begin{cases}
    q+r+s+t \leq d \\
    0 \leq q, r, s, t \leq d
    \end{cases}\Biggr\}
\end{equation}
where $p_1$ denotes the hypocentral depth, $p_2$ denotes the strike, etc.
We determine the coefficients contained in $\bm{W}$ and $\bm{B}$
by solving the linear system of equations
\begin{align*}
\bm{\Phi W}+\bm{\Psi B}&=\bm{A}_{\text{train}} \\
\bm{\Psi}^T\bm{B}&=\bm{0}
\end{align*}
where
\begin{equation}
    \bm{\Phi}=
    \begin{bmatrix}
    \varphi(\norm{\bm{p}_1-\bm{p}_1}) & \varphi(\norm{\bm{p}_2-\bm{p}_1}) & \ldots & \varphi(\norm{\bm{p}_{n_\text{train}}-\bm{p}_1}) \\
    \varphi(\norm{\bm{p}_1-\bm{p}_2}) & \varphi(\norm{\bm{p}_2-\bm{p}_2}) & \ldots & \varphi(\norm{\bm{p}_{n_\text{train}}-\bm{p}_2}) \\
    \vdots & \vdots & \ddots & \vdots \\
    \varphi(\norm{\bm{p}_1-\bm{p}_{n_{\text{train}}}}) & \varphi(\norm{\bm{p}_2-\bm{p}_{n_{\text{train}}}}) & \ldots & \varphi(\norm{\bm{p}_{n_{\text{train}}}-\bm{p}_{n_{\text{train}}}}) \\
    \end{bmatrix}
\end{equation}
and $\bm{\Psi}$ is the matrix of polynomials up to degree $d$.
For the RBFs $\varphi(\cdot)$, we test three types of kernels:
thin plate spline ($\varphi(r)=r^2\ln(r)$), cubic
($\varphi(r)=r^3$) and quintic ($\varphi(r)=-r^5$). For each kernel,
we solve the linear system using the minimum polynomial degree (equal to one
for the thin plate spline and cubic kernels, and degree
two for the quintic kernel).
Note that we do not apply smoothing, which implies
that the POD coefficients are exactly reproduced for the training dataset, i.e., for any $\bm p \in \bm P_\text{train}$.
The type of RBF kernel (i.e., thin plate spline, cubic or quintic)
is a hyperparameter that we tune using cross-validation.

\item
Artificial neural networks are another type of function approximator for the POD
coefficients \cite{hintonConnectionistLearningProcedures1990,hesthavenNonintrusiveReducedOrder2018b}.
We use a fully connected, feed-forward structure
in which the neural network consists of an input
layer, two hidden layers, and an output layer. This type of structure
is also known as a multilayer perceptron (MLP). In this type of model, the input layer receives the four earthquake source parameters
from $\bm{p}$, and the
output layer returns the predicted POD coefficients. Each neuron in the hidden layers receives an input from each neuron in the previous layer, evaluates an activation function (ReLU),
then passes the information forward through the network to every neuron in the next layer using determined weights and biases. We optimize the weights and biases by
training the neural networks with the Adam optimizer \cite{kingmaAdamMethodStochastic2017}
and the squared error loss function. Furthermore, we follow the method of \citeA{swischukProjectionbasedModelReduction2019}
to predict a subset of the POD coefficients which correspond to the
$r$ largest singular values, choosing to train the neural networks
using $r=$ 10, 30, and 50 modes from the SVD (Eq. \eqref{eq:svd}).
The hyperparameters for MLP consist of the rank $r$
and the number of neurons in each hidden layer.

\item
Random forests (RF) are a type of
ensemble estimator in which
many decision trees are constructed, and the final estimate
is the mean of each decision tree's individual estimate
\cite{breimanRandomForests2001}. We here use both 50 and 100 decision trees.
The hyperparameters are given by the number of decision trees.

\item
$k$-nearest neighbors (kNN) is a relatively
simple type of approximator in which the predicted value is given as the
mean of the $k$-nearest neighboring points. We apply uniform weighting
to each point in the neighborhood and use the $\ell_2$ norm as the distance
metric. We test using 3, 5, and 7 neighbors which define the hyperparameters.
\end{enumerate}

\subsubsection{Training procedure}
Before training the function approximators RBF, MLP, RF, and kNN, we standardize both the
inputs (earthquake source parameters, $\bm P$) and outputs (POD coefficients, $\bm A$) such that they
have a mean of zero and standard deviation of one.
Note that we use all available modes (i.e., all columns of $\bm{A}$) to train the ROMs,
except for the MLP approximator where we allow the number of modes
to be a hyperparameter. To determine the values of the hyperparameters,
we use $k$-fold cross-validation with $k=5$. In this procedure,
the training dataset is further subdivided into 5 folds. For each of the folds,
the remaining $k - 1$ folds are used to train the approximators, and the
fold that was not included is the validation dataset.
To measure the error of a ROM with a given set of hyperparameters, we
use both the mean absolute error and the mean absolute percentage error.
We define the local (in parameter space) mean absolute error as
\begin{equation}
    {MAE}(\bm p)  =\frac{1}{n_r}\norm{\bm{q}(\bm p)-\tilde{\bm{q}}(\bm p)}_1
\end{equation}
and the local mean absolute percentage error as
\begin{equation}
    {MAPE}(\bm p) =\frac{1}{n_r}\norm{\frac{\bm{q}(\bm p)-\tilde{\bm{q}}(\bm p) }{\bm{q}(\bm p) } }_1,
\end{equation}
where $n_r$ is the number of discrete points in space where the PGV map $\bm q$ is defined.
The average mean absolute error and average mean percentage error are obtained
by average the local metrics over the discretely sampled parameter space $\bm P_x$
and are defined via
\begin{align}
    \overline{MAE} (\bm P_x)  &=\frac{1}{n_x}\sum_{i=1}^{n_x}{MAE}{(\bm p_i)} \\
    \overline{MAPE} (\bm P_x) &=\frac{1}{n_x}\sum_{i=1}^{n_x}{MAPE}{(\bm p_i)},
\end{align}
where $\bm p_i \in \bm P_x$ and $n_x = \text{card}(\bm P_x)$ is the number
of parameter instances in the dataset $\bm P_x$.
When performing cross-validation, we select the hyperparameters
for each type of approximator that result in the lowest $\overline{MAE}$ averaged across all five folds. We then
re-train the approximators on the entire training dataset
using the cross-validated hyperparameters and evaluate the $\overline{MAE}$ and $\overline{MAPE}$,
now on the completely independent testing dataset. Altogether, this process
allows us to find the best-fitting ROMs while also giving us independent
estimates of the ROM errors.

\section{Results}
\label{sec:results}
\subsection{Information content of the proper orthogonal decomposition}
After generating the simulation data matrix $\bm Q$,
we first present the singular value distribution resulting from performing the SVD.
The efficiency and applicability of a reduced-order model for a given physical problem
is directly related to the singular value distribution.
Ideally, the singular values will rapidly decay, implying that $\bm Q$ has low-rank.
To characterize the singular value distribution of $\bm Q$ we use
the relative information content (RIC) given by
\begin{equation*}
    \text{RIC}(r)=\frac{\sum_{i=1}^r \sigma_i^2}{\sum_{i=1}^{n_s} \sigma_i^2},
\end{equation*}
where $r$ is the number of modes (left singular vectors).
Using $\text{RIC}(r)$, low-rank matrices will result in RIC $\approx 1.0$ when $r$ is small.
Furthermore, the RIC may be used to guide the
number of modes to use in defining the ROM \cite{swischukProjectionbasedModelReduction2019},
because the higher-order modes contribute little to the solution such that they may be
truncated.

\begin{figure}
    \centering
    \includegraphics[width=8cm]{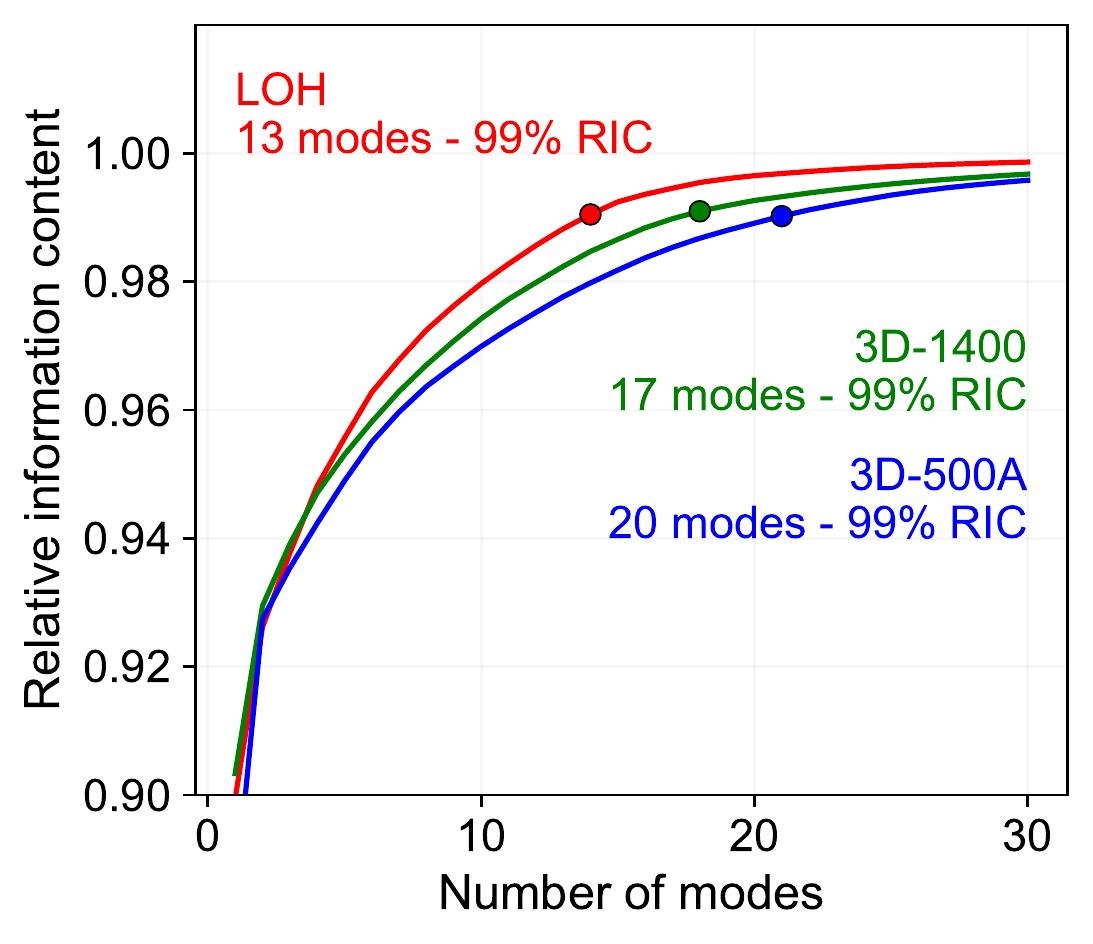}
    \caption{Relative information content (RIC) plotted as a function of the number of modes
        for the 5000 simulated PGV maps produced by the forward
        models (LOH in red, and \fm\ in green, \fmm\ in blue).
        Note that the horizontal axis is limited to show only the first 30
        modes. We illustrate this result in a movie  (Movie S1) that shows
        PGV maps defined using an increasing amount of $r$ left singular vectors, where
        we observe that the maps are well-approximated by the first few modes.}
    \label{fig:svals}
\end{figure}

In our generated datasets for all three forward models (LOH, \fm, and \fmm)
using $n_s = 5000$, we find that only a few modes are necessary to capture a significant
portion of the information content (Figure \ref{fig:svals}).
Specifically, only 13 modes are required to recover 99\% of the
information content for the data generated by LOH, 17 modes are
required for \fm, and 20 modes are required for \fmm. These results indicate that the simulated PGV maps,
independent of the forward model complexity, all have
low-rank. Since the number of modes reached to capture 99\% of the information within $\bm Q$
is only slightly larger for 3D velocity model setups, %
this suggests that the PGV-based ROM may be only slightly more
difficult to approximate (i.e., less accurate) compared to the PGV maps associated with LOH.

\subsection{Quantitative comparison of function approximators}
\begin{figure}
    \centering
    \includegraphics[width=12cm]{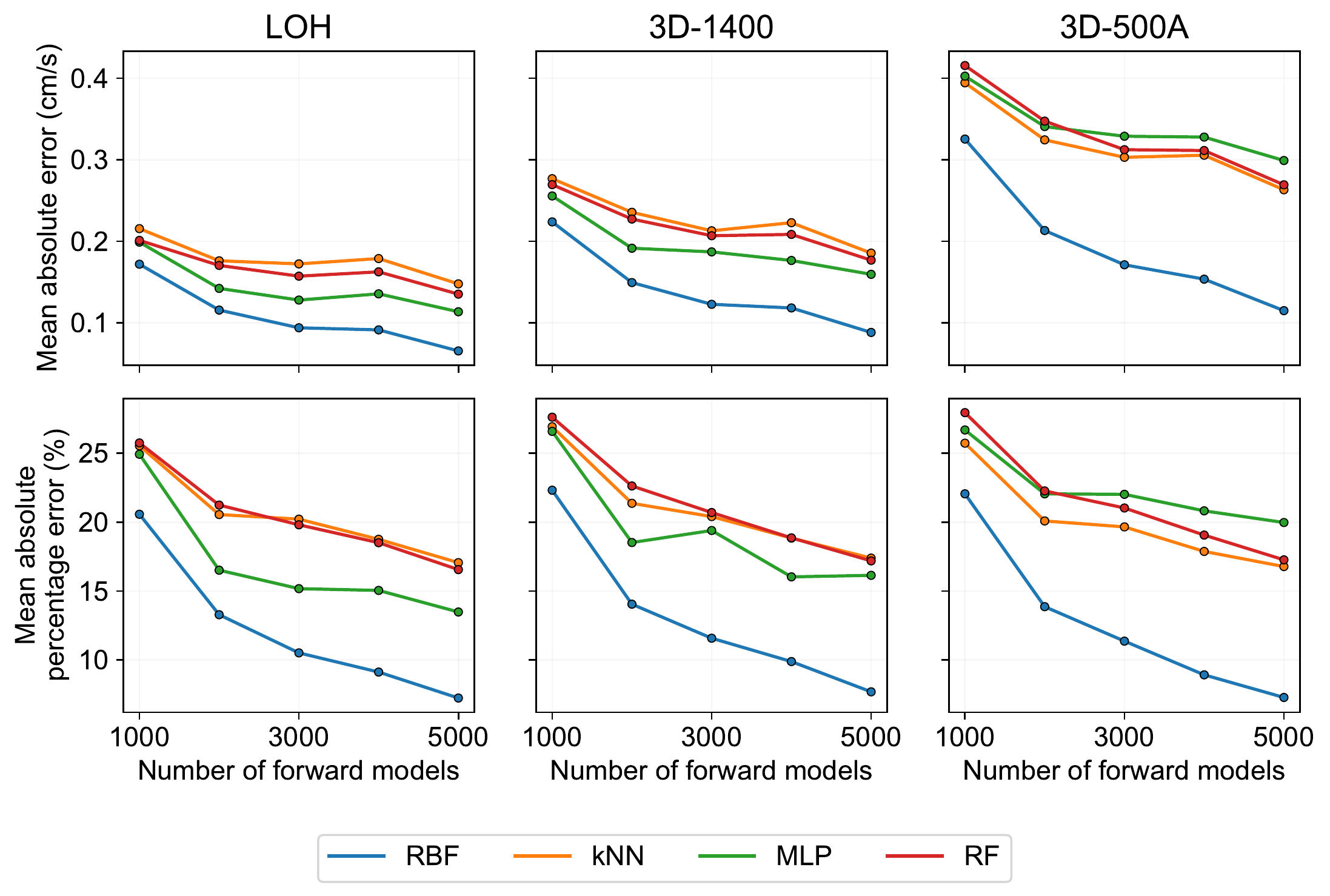}
    \caption{$\overline{MAE}$ (top row) and $\overline{MAPE}$ (bottom row) errors for the different
        ROMs using the LOH (left column) and \fm\ (right column)
        forward models. Each approximator is shown with a different
        colored solid line (see legend). For each number of forward models ($n_s$), we
        evaluate the errors using the testing dataset and the
        best hyperparameters selected from 5-fold cross-validation (see
        Table S2 for the selected hyperparameters). The mean values
        of $d_{\text{nearest}}$ for the testing datasets
        are 0.13, 0.11, 0.10, 0.09, and 0.09
        for the evaluations made using 1000, 2000, 3000, 4000, and
        5000 forward models, respectively.}
    \label{fig:interp-errs}
\end{figure}

Here, we present the results of training the ROMs using different
types of function approximators. In particular, we investigate the
errors between the ROM predictions and the simulated PGV maps in the testing
dataset. We examine how the ROM errors change depending on the
amount of training data. We train the ROMs with increments of $n_s/5$ forward models
and determine the cross-validated hyperparameters for each approximator
as described in Section \ref{sec:rom}. For each approximator, we show the errors
(both $\overline{MAE}$ and $\overline{MAPE}$) corresponding to the hyperparameters that result
in the lowest $\overline{MAE}$ (Figure \ref{fig:interp-errs}).
We indicate the hyperparameters that were selected for the final evaluation
of each approximator using $n_s$ forward models in Table S2.

Figure~\ref{fig:interp-errs} shows that all approximators
generally result in ROMs with lower error as more simulations
are added to the training dataset. We observe that the RBF-based ROM consistently has the lowest
error out of all ROMs that we consider, a result that is independent of the type of
velocity model (either 1D or 3D). When trained with the complete set of training data, the RBF-based ROM performs with
an average absolute PGV error of $\overline{MAE}=0.07$ cm/s for the LOH
forward model, $\overline{MAE}=0.09$ cm/s for \fm, and $\overline{MAE}=0.11$
cm/s for \fmm. The approximate values for $\overline{MAPE}$ are 7\%,
8\%, and 8\%, respectively.
The local slope of the error curves,
especially for $\overline{MAE}$ are variable as $n_s$ increases.
Specifically for $n_s = 3000$,  we observe that the local slope for $\overline{MAE}$ is approximately zero,
and for all but the RBF interpolator is actually positive.
This indicates that some points in the parameter space which are being sampled are %
not particularly informative for the ROM.
However, for higher values of $n_s$, the continuing downward slope
of $\overline{MAPE}$ for the RBF-based ROM suggests that
additional training data might further decrease the ROM errors.
Similar trends are observed for all approximators with LOH across both error metrics.
Additionally, the slope of the lines in going from $n_s = 4000$ to 5000 is approximately the
same across all approximators.
The same observations are observed for \fm, with the exception of the MLP approximator.
MLP exhibits a smaller slope for large values of $n_s$ in $\overline{MAE}$
and an approximately zero slope in $\overline{MAPE}$ when $n_s$ increases from 4000 to 5000.

\begin{figure}
    \centering
    \includegraphics[width=\textwidth]{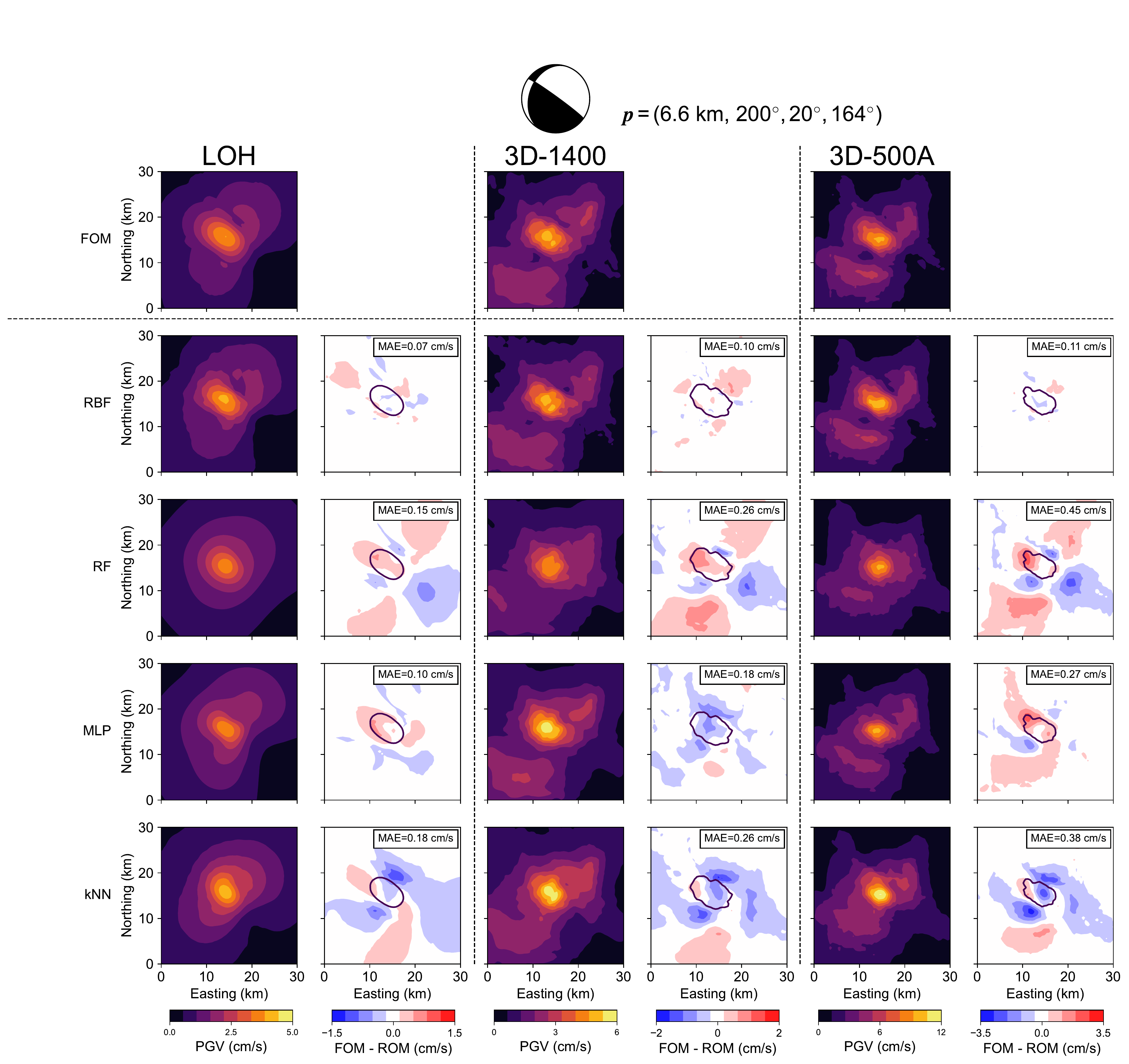}
    \caption{Comparison of the ROM solutions using different
    function approximators for a single earthquake source (focal mechanism shown
    in the first row). The second row shows the target PGV maps for
    LOH (left) and \fm\ (right). The subsequent panels show the
    ROM predictions and errors, where each row is for a different
    approximator (see row labels at left). The approximators are sorted
    from lowest to highest error (top to bottom), and each error map is indicated
    with the $MAE$ in the top right corner. The error maps outline
    the areas of strongest shaking where the PGV is
    90\% of the largest PGV produced by each FOM.}
    \label{fig:interp-preds}
\end{figure}

In addition to examining the overall errors on the testing dataset,
it is informative to examine spatial errors within the PGV maps.
We show examples of simulated PGV maps
created using all three velocity models, the predictions from
the four types of approximators, and map-views of the errors (Figure \ref{fig:interp-preds}).
The map views of the errors illustrate locations in which the
ROM produced accurate PGV estimates, or where the PGV was
over/under-predicted.
For the specific earthquake source in this example with oblique slip,
the PGV is strongest a few km northwest of the epicenter, with simulated PGVs
reaching up to 12.0 cm/s for the \fmm\ setup.
Averaged across all receivers,
the RBF ROM gives the smallest error ($MAE$=0.07 cm/s for LOH),
while the kNN ROM gives the largest error ($MAE$=0.18 cm/s).

Although the errors show significant differences depending
on the type of function approximator,
each approximator generally reproduces the overall spatial distribution
and amplitudes of the forward model ground motion map. The RBF interpolator
does significantly better in correctly predicting ground motions in the areas of strongest
and weakest shaking. Notably, the errors for the RBF ROM
are near zero where the shaking is the strongest.
The other approximators share some common problems,
such as over-prediction at the eastern edge of the domain and
under-prediction near the southwest.
The kNN, in particular, tends to overestimate the PGVs
in the near-epicentral area of strong shaking. Overall,
we conclude that the RBF ROM performs exceptionally well at recovering the
spatial distribution of ground motion compared to the other approximators.

\subsection{ROM solutions for different sources}
\begin{figure}
    \centering
    \includegraphics[width=\textwidth]{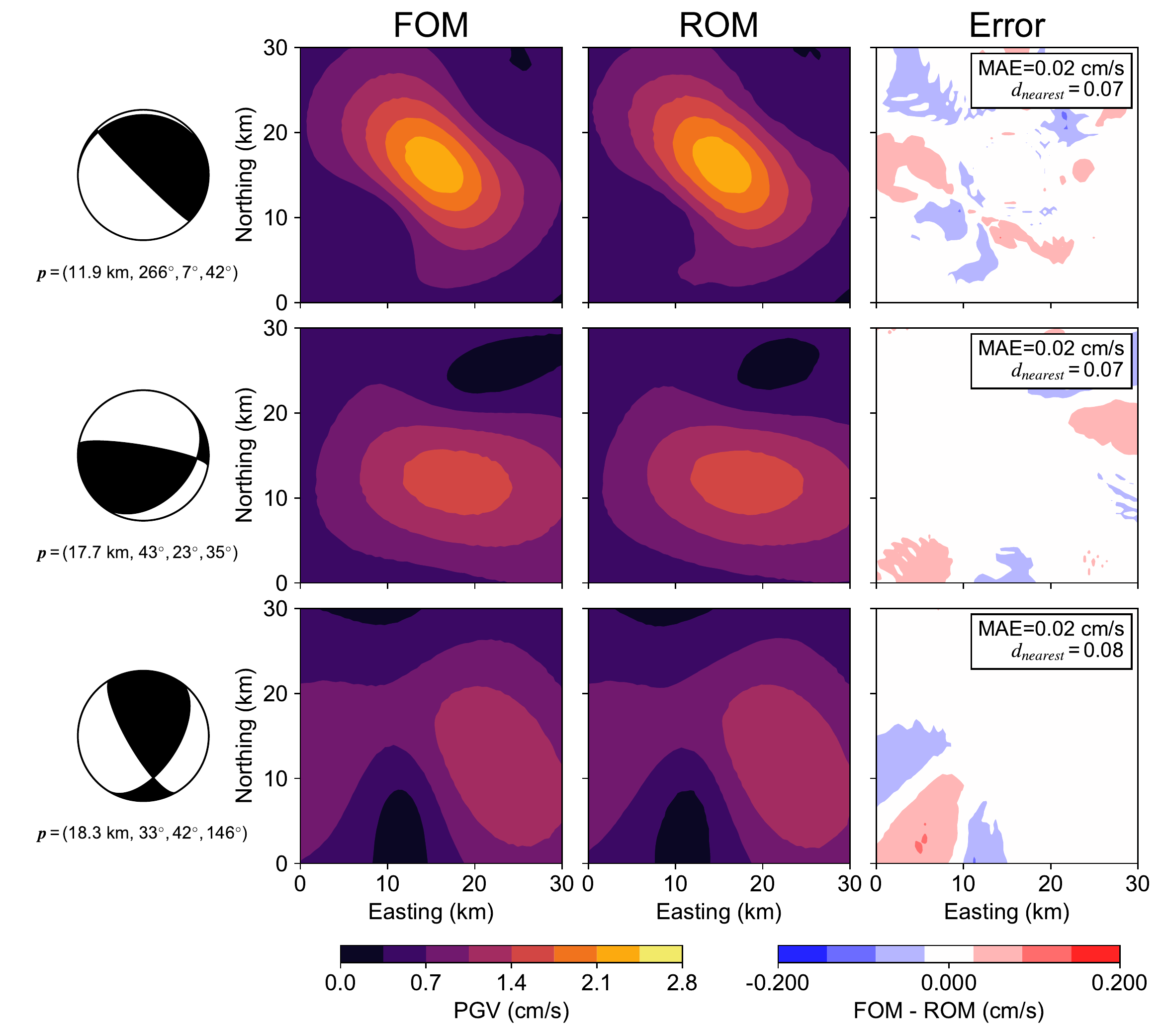}
    \caption{Plots of three simulated PGV maps (second column), RBF ROM predictions (third column), and errors (fourth column)
    for the LOH forward model. Each row indicates the results for
    a different earthquake source in the testing dataset, where the source
    parameters are indicated by the focal mechanisms in the first column.
    The error maps show the difference between the FOM and ROM solutions,
    and indicate the mean absolute error (MAE) and distance to the nearest
    training location ($d_{\text{nearest}}$). The complete set
    of errors on the testing data can be viewed as histograms in Figures S4
    and S5.}
    \label{fig:loh1-sources}
\end{figure}

\begin{figure}
    \centering
    \includegraphics[width=\textwidth]{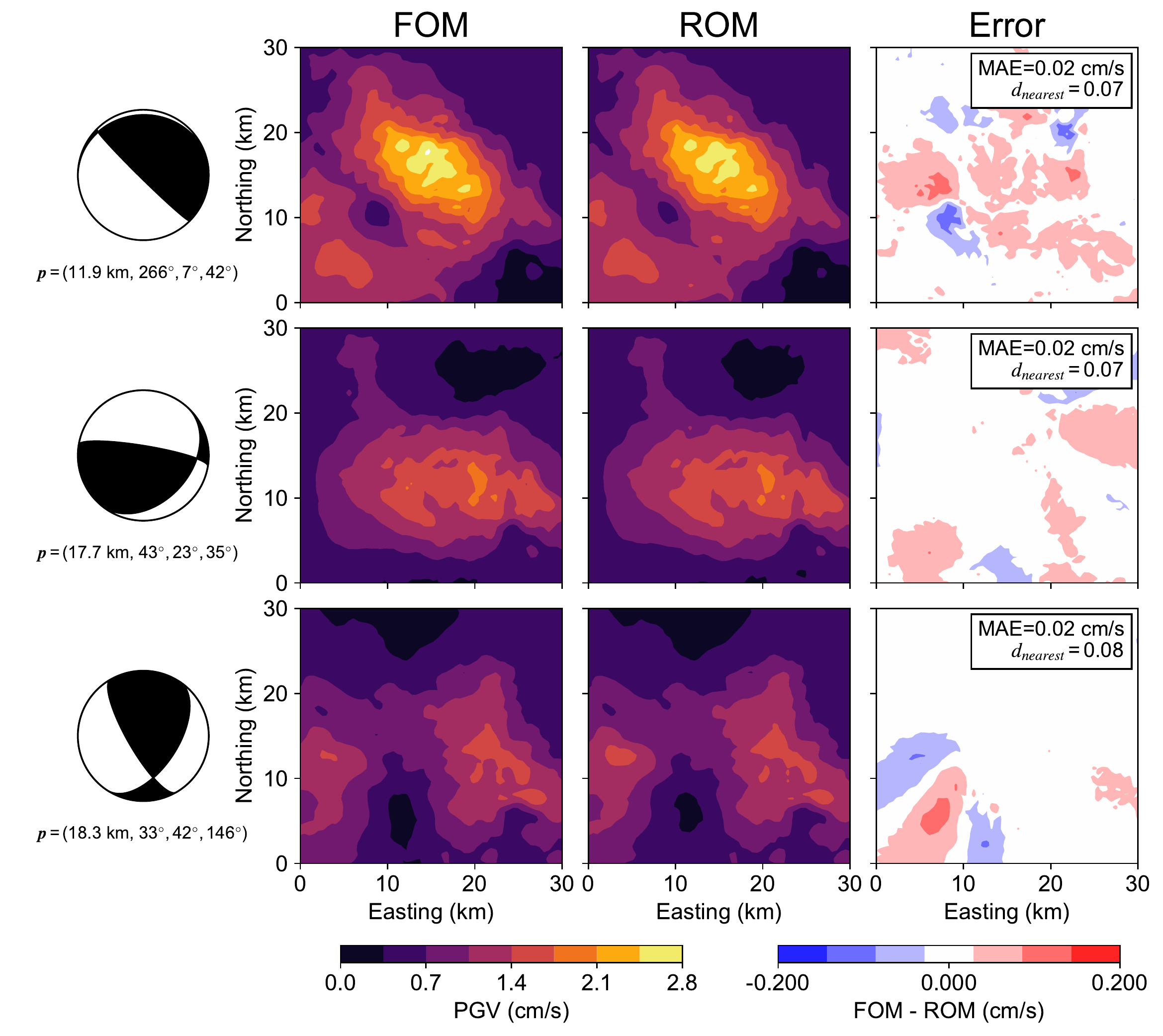}
    \caption{Plots of the simulated PGV maps, ROM predictions, and ROM errors
    for the same sources as shown in Figure \ref{fig:loh1-sources}, but shown here for the \fm\ forward model.
    The forward model includes 3D velocity structure (with minimum S-wave speeds of
    1400 m/s) and topography.} %
    \label{fig:3d-sources}
\end{figure}

\begin{figure}
    \centering
    \includegraphics[width=\textwidth]{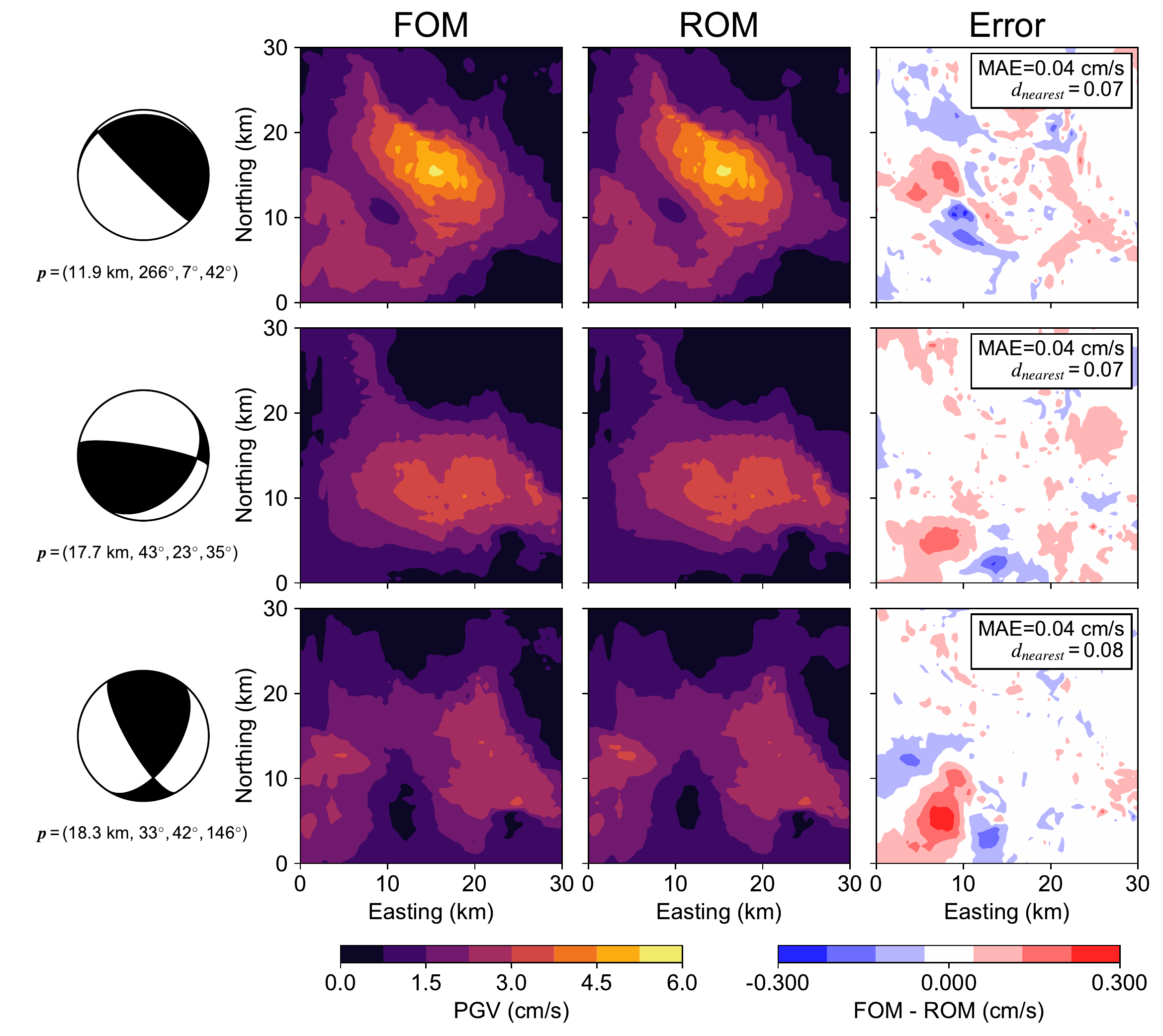}
    \caption{Plots of the simulated PGV maps, ROM predictions, and ROM errors
    for the same sources as shown in Figure \ref{fig:loh1-sources}, but shown here for the \fmm\ forward model.
    The forward model includes 3D velocity structure (with minimum S-wave speeds of
    500 m/s), topography, and 3D variable viscoelastic attenuation.} %
       \label{fig:3d-500a-sources}
\end{figure}

We next demonstrate the ROM's ability to predict the changes in the
spatial patterns and amplitudes of peak ground motions
when the hypocentral depth and focal mechanism of the earthquake source are
subject to change. We show examples of PGV maps created with the LOH velocity model using
three different sources from the independent testing dataset (Figure \ref{fig:loh1-sources}).
The PGV maps that result from the same sources for the \fm\ forward model are shown in Figure \ref{fig:3d-sources},
and the maps for \fmm\ are shown in Figure \ref{fig:3d-500a-sources}.
In these figures, we select three sources in $\bm{P}_{\text{test}}$ where
the distance $d_{nearest}$ is close to the expected value (0.09)
of the distribution of distances for any source in $\mathcal{P}$. We simplify the presentation
by only showing the RBF interpolator results, as this is our
preferred function approximator for the ROM.
As the source focal mechanism changes, we see the ROM correctly
reproduces the changes that occur to the PGV map.

We see that the physically more complex forward models \fm\ and \fmm\ produce more complex spatial PGV distributions,
compared to the much smoother distributions produced by the simpler LOH models.
Comparing Figures \ref{fig:loh1-sources} \& \ref{fig:3d-sources}, the ROM
for \fm\ has only slightly more difficulty in predicting the smaller-scale
features in the \fm\ data, and the errors
are only slightly larger compared to LOH. These small differences between LOH and \fm\ are
consistent with the overall trends from the testing dataset. Comparing
\fm\ against \fmm, we see that \fmm\ results in larger PGVs due to the
slower wave speeds included in the velocity model. For these particular sources,
the MAE values are about 2.0 times larger for \fm\ compared to \fmm.
The complete set of error measurements on the testing dataset
are shown as histograms in Figures S4
and S5, respectively, for the three forward modeling setups.

\begin{figure}
    \centering
    \includegraphics[width=\textwidth]{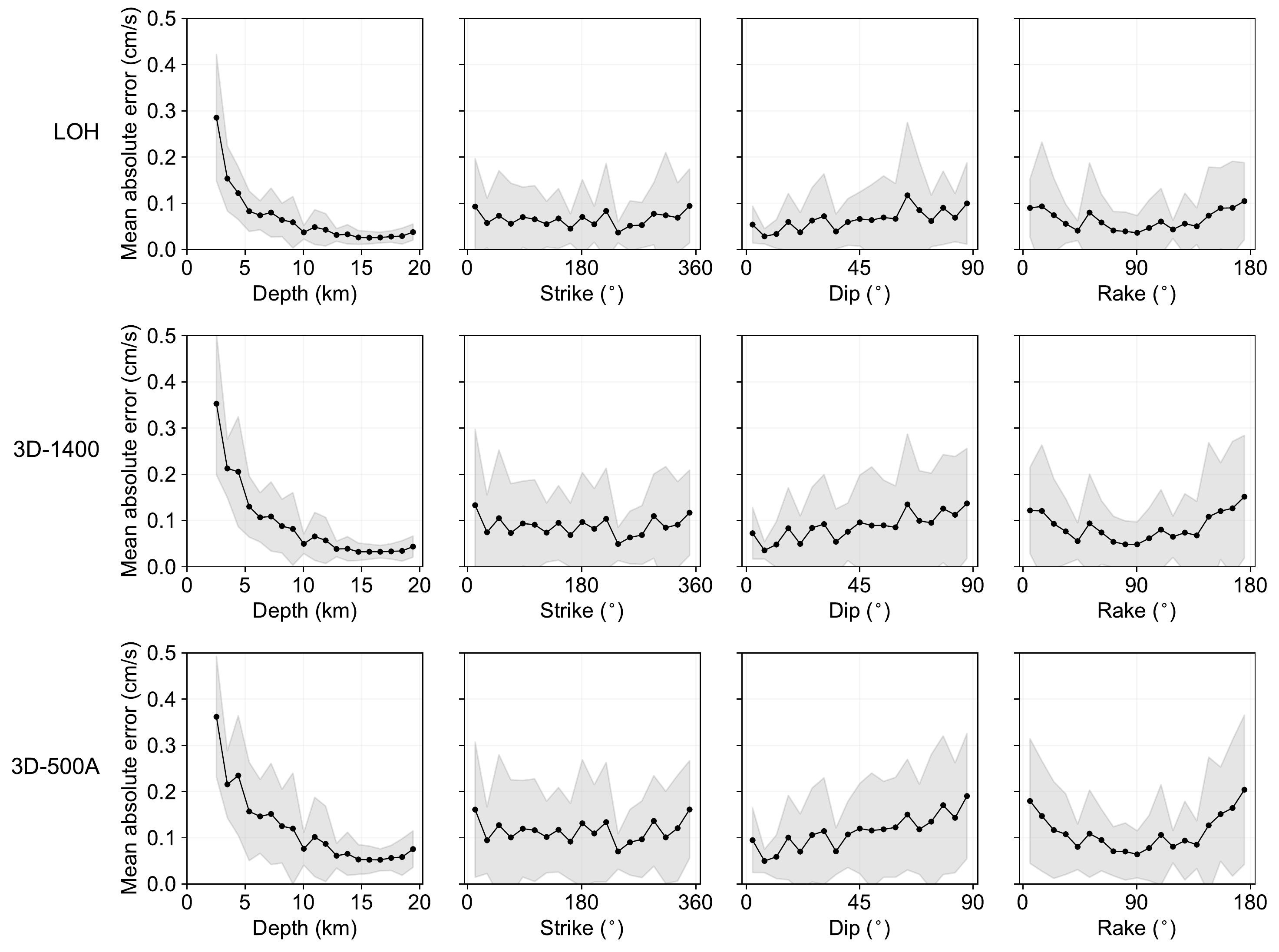}
    \caption{Mean absolute errors for the PGV predictions on the testing dataset, plotted as a function of each
    earthquake source parameter. LOH predictions are shown in the top
    row, and \fm predictions in the bottom row. For each parameter, we compute the
    mean (black line and dots) in 20 different evenly spaced bins and plot $\pm$ 1.0
    standard deviation of the error in each bin (shaded area).}
    \label{fig:errs-params}
\end{figure}

We inspect the accuracy of the ROMs for
different values of the input source parameters. In Figure \ref{fig:errs-params},
we show the average and standard deviation
of $MAE$ for 20 evenly spaced bins for each
of the four source parameters.
For all three forward models, we observe that the shallower events show larger errors compared to deeper events.
This is partially explained by the larger
PGVs expected for shallow sources, although the relative $MAPE$ errors
show a non-negligible correlation with hypocentral depth (Figure S7). %
There is more variability in the errors %
for shallow events compared to deep events as indicated by the
standard deviations.
Shallow sources introduce additional near-source effects to ground shaking which may be underrepresented in the training dataset following using our uniform sampling strategy.
For the other source parameters related to geometry (i.e., focal mechanism, parameterized as strike, dip, and rake), the errors do not show significant
correlations. The standard deviations in each bin are larger
for these parameters, indicating that the ROM errors
are primarily controlled by the source hypocentral depth. Overall,
these results imply that the ROM errors are less dependent on the focal mechanism
than on the source hypocentral depth.

\subsection{Relationship between ROM errors and distance to training locations}

It is reasonable to expect that the ROM errors will be smaller for the testing parameters
that have a closer neighbor in the training dataset (i.e.,
a smaller value for $d_{\text{nearest}}$). We indeed see this
trend for our testing dataset
(Figure S2) where the errors correlate with the
distance to the nearest neighbor. The largest error of $MAE_i\approx 0.8$ cm/s
(for \fm) occurs when the distance between the testing parameter and the
nearest training parameter is approximately equal to the
fill distance of 0.17.
When the Euclidean distance
$d_{\text{nearest}}$ increases beyond 0.12 units, the error becomes
significantly larger indicating that the interpolation quality of the ROM
degrades when the distance to the nearest training point is too large.
This relationship between $d_{\text{nearest}}$ and
ROM error can be used to estimate ROM errors
for earthquake sources contained in $\mathcal{P}$ where
we do not have a FOM solution. For example, for an earthquake
source with a parameter $\bm{p}$ that is 0.1 distance units away from the nearest
training FOM location, we expect the $MAE_i$ of the ROM-predicted
PGV map for \fm\ to be, on average, 0.1 cm/s.

To quantify the value added by the ROM, we examine the predicted
PGV maps that would result from a simple approach of selecting the nearest-neighboring
training FOM solution. As shown in Figure S3,
the error for all forward models is about 2.5 times larger when selecting the
nearest training FOM solution, compared to our preferred ROM estimates.
This result illustrates that our ROMs are successfully interpolating
between the training FOM solutions to obtain better estimates than could
be obtained from just the simulated data.

\subsection{Quantification of the effect of depth uncertainty on PGV predictions} %
We now present an exemplary application using the ROMs to
study ground motion effects related to variable earthquake source parameters.
In current operational earthquake early warning algorithms, the
source focal depths are typically fixed at 8.0 or 10.0 km
to minimize computational costs \cite{boseFinDerImprovedRealtime2018b,chungOptimizingEarthquakeEarly2019}.
These depth assumptions are reasonable in California where most earthquakes occur
on shallow crustal faults \cite{brown2011development} but may negatively
affect alerting accuracy in other regions, for example, intraslab earthquakes in the Pacific
Northwest \cite{thompsonEffectFixingEarthquake2022}.
With the ROMs, we can quantify the potential error that
may be committed by an incorrect assumption of the hypocentral depth. We use our
\fmm\ ROM to make predictions for 1 million different focal mechanisms,
obtaining the PGV maps for both a shallow (2.0 km) and a deep (8.0 km) source.
The focal mechanisms are obtained by sampling a uniform grid
of 100 points along each of the three focal mechanism parameters.
Using the RBF ROM to create ground motion maps for these 2 million earthquakes
requires $\sim$1320 s on a personal laptop\footnote{MacBook Pro (2020), Intel(R) Core(TM) i5-1038NG7 CPU @ 2.00GHz}.
This translates to a time-to-solution for a single earthquake scenario of $\sim$0.66 ms, or $\sim 0.6 \times 10^{-6}$ CPUh\footnote{As measured by the macOS application ``Activity Monitor''}.

\begin{figure}
    \centering
    \includegraphics[width=\textwidth]{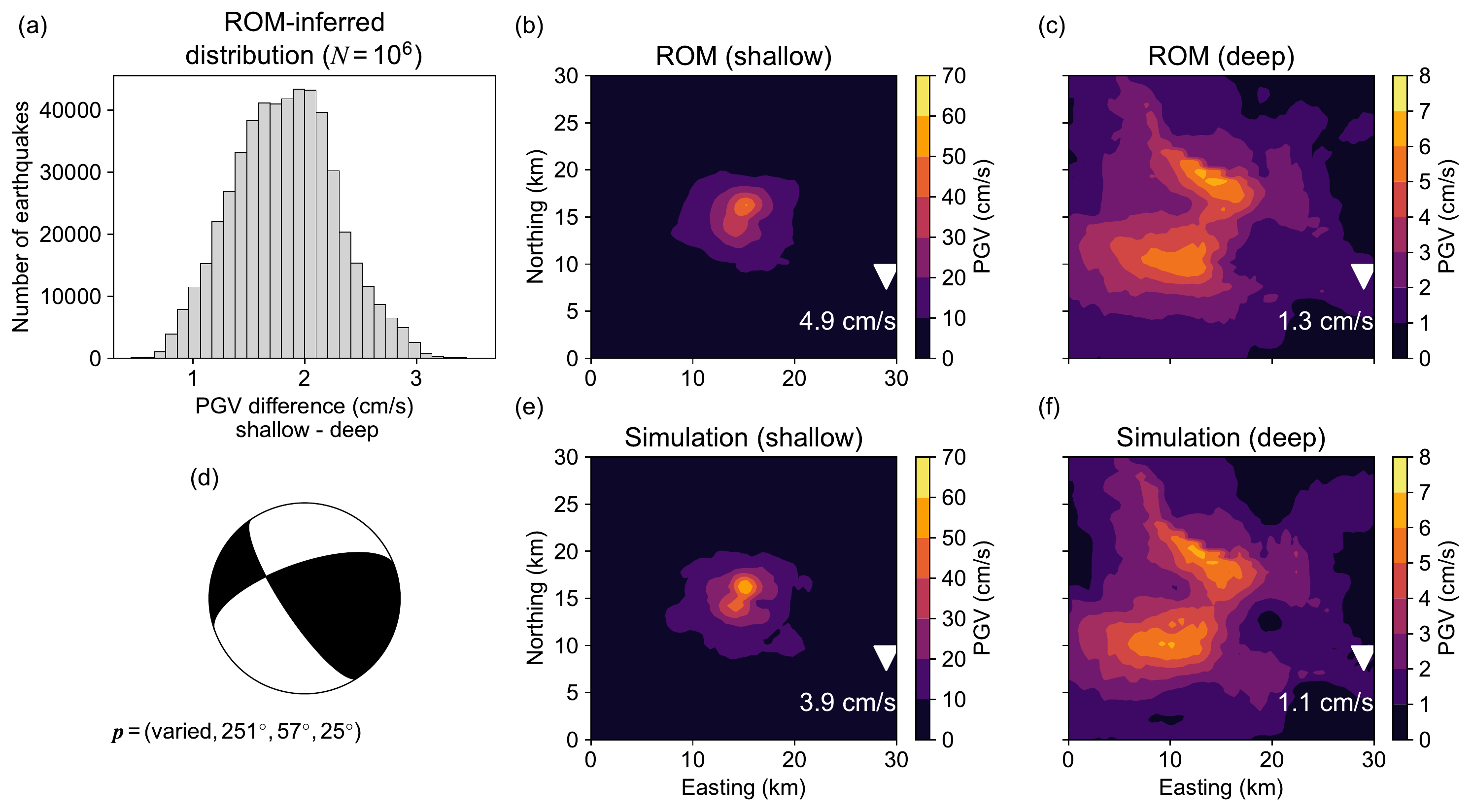}
    \caption{Histogram showing the distribution of differences between the shallow (2.0 km)
    and deep (8.0 km) source PGV predictions for 1 million different
    focal mechanisms (a). The ROM PGV predictions
    for the worst-case focal mechanism (d) are shown
    for the shallow source (b) and deep source (c).
    The corresponding simulations are shown in
    the bottom row in (e) and (f). In the PGV maps, the site of interest
    is indicated by a white triangle, and the site's PGV value
    is printed.}
    \label{fig:exp1}
\end{figure}

To simplify our analysis of ground motion variability related to the
source parameters, we chose to focus on an individual site
in our computational domain. We select the Clearwater Power Plant (33.891 N, 117.609 W) near the
southeastern part of our domain (see Figure \ref{fig:forward}a). At this site,
the ROM predictions show that the PGV varies significantly depending on the focal mechanism (Figure \ref{fig:exp1}a).
Averaging across all focal mechanisms, the mean difference in PGV between the shallow
and deep sources is 1.8 cm/s; however, the maximum PGV difference
is 3.6 cm/s for certain focal mechanisms. This result implies that, in the worst-case scenario, the
PGV predictions based on an 8.0 km hypocentral depth could underestimate the true PGV
at this site by up to 3.6 cm/s if the true hypocentral depth is 2.0 km.

One advantage of the ROM approach is that we can deterministically verify individual
estimates by performing additional forward simulations. Here,
we verify the ROM-based estimate for the worst-case scenario
by performing two additional \fmm\ simulations
(see Section \ref{sec:forward}) using the focal mechanism that resulted in the largest ROM-predicted
PGV difference (between shallow and deep sources) at our site of interest.
In the forward simulations, the PGV difference
between the shallow and deep sources is 2.8 cm/s for the site.
In comparison, the ROM-based estimate of the PGV difference is 3.6 cm/s.
The discrepancy likely results from the ROM over-estimating the PGV for
the shallow source. However, this amount of error is well within the expected error
as indicated by the mean absolute errors from the testing dataset at
this site (Figure S6). The amount of error for the ROM-estimated
PGV differences (between shallow and deep sources) that are closer to the
mean will be similar to the worst-case scenario because the errors are
uncorrelated with the focal mechanism angles (Figure \ref{fig:errs-params}).
Overall, this experiment highlights the potential in using ROMs to identify
potentially damaging ground motions that might be overlooked when
using a smaller sample size of physics-based simulations.

\subsection{Sensitivity analysis using Pearson correlation
coefficients} %
Finally, we analyze the correlations between variable earthquake
source parameters and the PGV predictions at the same site of interest.
This type of analysis allows us to quantify the relative importance of the model inputs
(earthquake source parameters) with respect to the model output (PGV) \cite{homma1996importance}.
Here, we perform a global sensitivity analysis, which is closely
related to the problem of forward uncertainty quantification \cite<e.g.,>{pagani2021enabling}.

\begin{figure}
    \centering
    \includegraphics[width=\textwidth]{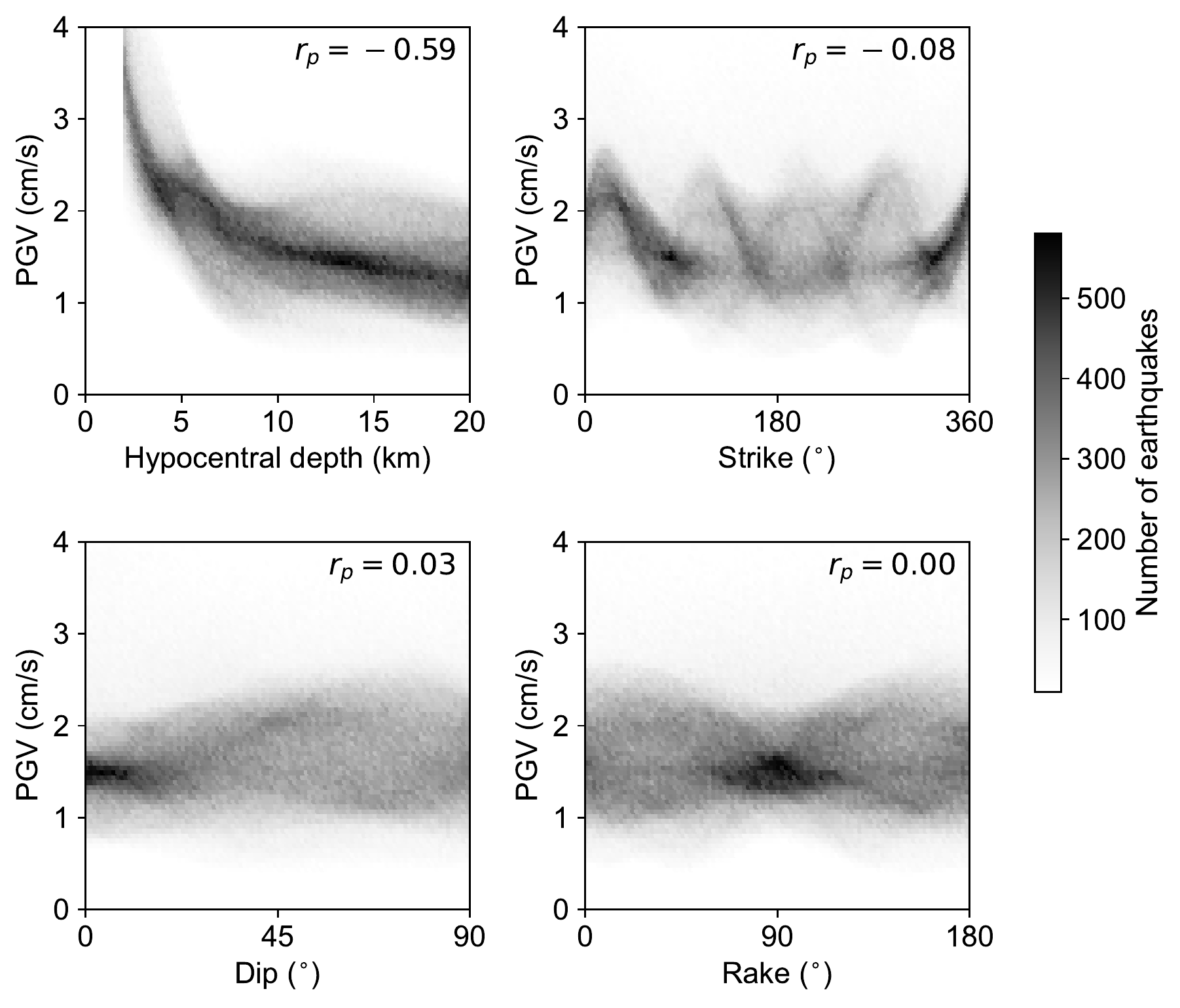}
    \caption{2D density plots of the \fmm\ PGV predictions at the site of interest, plotted
    against each source parameter. In each bin, the number of earthquakes
    is indicated by the color bar to the right. The Pearson
    correlation coefficient is indicated in the top right of each panel.}
    \label{fig:corr}
\end{figure}

To measure the correlations, we again sample 1 million earthquakes,
where the source parameters are randomly sampled from uniform distributions.
After obtaining the ROM solutions at the Clearwater Power Plant for these sources,
we measure the correlation between each source parameter and the PGV
data using the Pearson correlation coefficient ($r_p$) \cite{pearson1895vii}.
We find a significant inverse correlation ($r_p=-0.65$)
between the PGV predictions and hypocentral depth (Figure \ref{fig:corr}). For the focal mechanism
angles, the correlation coefficients are near zero, implying that
there is no consistent linear relationship between any of the focal mechanism parameters
and PGV. However, there are features in the relationships between
source parameters and PGV that cannot be adequately
captured by the correlation coefficient. For example, the strike
clearly affects PGV at this site by means of radiation pattern effects which
cause the visible streaking patterns. Additionally, the kink in the
PGV scaling with depth at about 5 km affects the correlation coefficient
measurement. We note that these general correlations could also be measured
from the PGV ground motions produced directly by the FOM (Figure S9), while the
1 million ROM predictions provide finer detail compared to solely examining the
5000 FOM measurements. Overall, while the PGV at a site of interest may
vary significantly depending on the source parameters, the hypocentral
depth is the most significant contributor to the PGV variability
in our dataset.
This finding agrees with the accuracy analysis presented in previous sections.

Lastly, we quantify our site of interest's variability in PGV that results from
changing the source focal mechanism. We consider this type of variability to be analogous to the
inter-event terms that are often reported by GMMs which use mixed-effects
regressions, such as ASK14 \cite{abrahamson2014summary}. We bin our PGV data shown
in Figure \ref{fig:corr} into 30 equally sized hypocentral depth bins and compute
the mean and standard deviation for each bin. We find that the standard
deviation increases for increasing hypocentral depths, and ranges
from 0.18 to 0.32 natural log units (Figure S8). These variability measurements
are somewhat lower compared to GMMs, for example ASK14, which reports an inter-event
standard deviation of 0.38 natural log units for $M_w=5.4$. This is an
expected result, however, as the ROM-predicted PGV variability only accounts
for variations in focal mechanism and not other source properties (e.g.,
stress drop) which increase the inter-event standard deviations in GMMs.
We discuss this topic further in the following section.

\section{Discussion}
\label{sec:discuss}
We have constructed ROMs with the goal of predicting PGV maps
for three different model setups of increasing complexity. We indeed find that the ROMs
accurately predict PGVs (about 0.1 cm/s average error) for both the
1D layer-over-halfspace (LOH) forward model and the 3D velocity models
with topography (\fm\ and \fmm) where the ROM errors are only
slightly larger for the 3D velocity models. We find that the ROM approach works well for both linear elastic and
viscoelastic forward models of seismic wave propagation, including low S-wave velocities,
as shown by our results for the \fmm\ forward model.
The ROM ground motion errors are not uniformly distributed across the parameter space, however,
as the ROM errors tend to be larger (in both an absolute and relative sense)
for shallow events compared to deep events. While this study uses a low-discrepancy
Halton sequence to determine the FOM training data that inform the ROM, it may be
beneficial in future ROMs to non-uniformly sample the parameter space. For example,
we suspect the larger errors for shallow sources may result from the
slower near-surface velocities which cause
fewer training samples to be drawn per wavelength. As such, future ROMs
might benefit from drawing the training FOM locations
proportionally to the slowness in the velocity model, which would result
in more training data for shallow sources compared to deep sources.
The error for shallow sources could also be corrected using
iterative ROM assembly \cite<e.g.,>{lye2021iterative}
where the errors on the testing dataset
are used to guide which new
parameters will be sampled for the next iteration.

Future ROMs may investigate other function approximators beyond the four
types we considered here. We find RBF interpolation to perform the best
for our application of interest, while other studies have obtained good
results using more generalized Gaussian process regression
\cite{guoReducedOrderModeling2018}.
This approach may allow for more flexibility in the
kernel length scales which can be optimized to better fit the data,
as our RBF interpolation is essentially
a special case of Gaussian process regression.

Using a Chino Hills earthquake scenario as our target simulation,
we show that ROMs can be used as on-demand ground motion simulators
for modeling wave propagation up to 1 Hz.
This frequency target is consistent with the current
understanding that PGV is controlled by frequencies around 1 Hz
\cite<e.g.,>{rodgersRegionalScale3D2020}. However, the ROM predictions may
under-estimate real PGV for smaller earthquakes which are enriched in higher
frequency content. In addition, the minimum chosen $V_S$ in our
forward simulations does not account for materials with even slower wave speeds that
cause ground motion amplification in specific regions, for example, the San Francisco Bay area muds
\cite<e.g.,>{hough1990sediment}.
To generate rapid, physics-informed PGV maps targeted in these
regions for earthquake engineering and seismic hazard applications,
the amount of computing time spent in the FOM data
collection phase would increase due to the higher cost of each forward simulation.
For example, halving the minimum shear wave velocity (or doubling the resolved frequency)
would require a factor of $2^4=16$ higher computational cost (assuming
spatially adaptive unstructured meshes and local time-stepping).
However, we expect the ROM approach to be able to capture the
changes to the PGV maps, as we show that there it
little difference in the rank of the simulated
data matrices when we lower $V_{S,\text{min}}$ from 1400 to 500 m/s.
We suspect this would also be true
if the simulations resolved higher frequencies and accounted
for even slower near-surface wave speeds.
ROM-based ground motion maps from higher-frequency and physically more complex simulations
(e.g., including approximations of linear and nonlinear site effects), may help to better
understand source, path and site contributions to ground shaking
\cite<e.g.,>{hartzell1997variability,frankel2002nonlinear,hashashViscousDampingFormulation2002}.

While the aim of this study is to demonstrate how the ROM approach can emulate
a ground motion simulation scenario with variable source properties, we note in the following possible future extensions
that may generalize the ROMs for
ground motion prediction of real earthquakes with arbitrary source
locations and magnitudes.
For one, we expect the ROM approach will be applicable to forward models involving more
realistic earthquake source representations, such as finite kinematic or
dynamic rupture sources. In this study, we did not parameterize forward simulations with variable moment magnitudes.
While our PGV maps could be scaled to represent different $M_w$ using a magnitude-scaling term from GMMs, for example \cite{baltay2014understanding},
the point-source approximation we use here omits finite source effects such as directivity \cite{spudich2014comparison}.
We note that our simulations only consider a single epicentral
location, and will not accurately predict the path effects in regions
with different velocity structures. This issue may be investigated
by extending $\mathcal{P}$ to include latitude and longitude
as additional source parameters, and we expect the ROM approach to be useful
in that case to study basin or topographic effects that depend on the source-to-site azimuth
\cite<e.g.,>[]{thompsonBasinAmplificationEffects2020,stone2022topographic}.
Variations to our fixed source-time function would allow us to study variable stress drop,
which has been shown to correlate with peak ground acceleration in observational
datasets \cite{trugmanStrongCorrelationStress2018}.
Recorded ground motion observations are typically not available at
the required density to build ROMs. While we here rely on physics-based
simulations to obtain PGV maps, we expect that future work may
incorporate sparse, observed ground motions into the ROM framework
to provide additional constraints on the ground motion amplitudes
and variability.

While finite earthquake source effects on PGVs
may be readily included in ROMs with additional source parameters,
an alternative approach is to predict Green's functions instead of
PGV maps. This would allow the ground motion intensities to be computed
for arbitrary source-time functions, and allow for predictions using
finite-fault sources using the superposition property of the elastic
wave equation. However, this approach might be challenging
if the Green's functions do not possess low-rank structure
like the PGV maps we predict in this study. For this application,
deep learning and neural operator methods
for time-dependent output quantities may be considered
\cite<e.g.,>[]{yangSeismicWavePropagation2021a,yang2023rapid}
though they have yet to be applied to 3D
elastodynamic wave propagation in models that include topography. Future
directions in instantaneous ground motion estimates may involve
non-source-based approaches that use ROMs to forecast wavefield snapshots,
either used simulated or observed data \cite<e.g.,>{nagata2023seismic}. %

\section{Conclusion}
We develop a reduced-order modeling (ROM) approach and use it to %
instantly create peak ground velocity (PGV) maps that closely match
computationally expensive wave propagation simulations
with variable source depth and focal mechanisms.
We show that the simulated PGV maps have low-rank
structure which motivates our use of the interpolated proper orthogonal decomposition
technique. The accuracies of the ROMs are similar when a 1D or 3D-velocity model (including
topography, viscoelastic attenuation, and slow S-wave speeds) is used for the simulations of seismic wave propagation. %
We find %
that the ROM errors are primarily controlled by the hypocentral depth
(larger errors for shallow sources).
After comparing different approximation strategies with cross-validation,
we find that radial basis function interpolation performs well, showing
small (mean absolute PGV error $<0.1$ cm/s) errors on an independent testing dataset.
We lastly demonstrate the potential of using
ROMs to answer questions about ground motion variability that
would ordinarily be computationally intractable when using full complexity forward simulations of wave propagation.
We conclude that the
reduced-order modeling approach defined via the interpolated proper orthogonal decomposition
is a viable technique to obtain physics-informed estimates
of the relationship between source properties and PGVs
for ground motion simulation scenarios.

\section{Open Research}
\subsection{Data Availability Statement}
We use the Scikit-learn \cite{pedregosaScikitlearnMachineLearning2011}
and SciPy \cite{virtanenSciPyFundamentalAlgorithms2020}
Python modules for their implementations of the approximators used
in this study. We created the figures for this work using matplotlib \cite{hunterMatplotlib2DGraphics2007},
cartopy \cite{metofficeCartopyCartographicPython2010}, and obspy \cite{beyreutherObsPyPythonToolbox2010}.
We provide the simulated PGV maps that we used for training the ROMs
at the public GitHub repository for this paper \cite{rekoske_john_2023_8118300}.
This repository provides a Jupyter notebook which demonstrates using the iPOD
reduced-order modeling technique to create ROMs for predicting new
PGV maps.

\acknowledgments
We would like to thank the Editor Rachel Abercrombie and the Associate Editor
for their comments that helped to significantly improve this manuscript.
We also greatly appreciate a review from Jack Muir who helped
us improve the quality of the forward simulations and
clarify our methodology,
as well as an anonymous reviewer whose feedback helped us improve
our presentation of the ROM error analysis and clarify the benefits of our
reduced-order model.
This material is based upon work supported by the National Science Foundation
Graduate Research Fellowship Program under Grant No. DGE-2038238.
Any opinions, findings, and conclusions or recommendations expressed in this
material are those of the author(s) and do not necessarily reflect
the views of the National Science Foundation.
We gratefully acknowledge the Gauss Centre for Supercomputing e.V. (\url{www.gauss-centre.eu})
for providing computing time on the GCS Supercomputer SuperMUC-NG at
Leibniz Supercomputing Centre (\url{www.lrz.de}), in project pn49ha. %
We acknowledge additional support from the National Science Foundation (NSF Grant No. EAR-2121568),
the National Aeronautics and Space Administration (80NSSC20K0495),
from the European Union's Horizon 2020 Research and Innovation Programme (TEAR ERC Starting,
grant No. 852992), Horizon Europe (DT-GEO,
grant No. 101058129, Geo-INQUIRE, grant No.101058518, and ChEESE-2P, grant No. 101093038), %
and the Southern California Earthquake Center (SCEC award 22135).

\bibliography{rom_pgv.bib}
\end{document}